\documentclass[a4paper,11pt]{article}
\pdfoutput=1 
\usepackage{jcappub} 

\usepackage[T1]{fontenc} 

\usepackage{mathalfa}
\usepackage{graphicx}
\usepackage{caption}
\usepackage{subcaption}
\usepackage{amsmath,amssymb}
\usepackage[font=small,labelfont=bf]{caption}
\usepackage{verbatim}
\usepackage{placeins}
\usepackage{tikz}
\usetikzlibrary{arrows, positioning, angles,quotes, calc, intersections}

\def\mean#1{\left< #1 \right>}

\usepackage{color}

\usepackage[normalem]{ulem}

\title{Optimal galaxy survey for detecting the dipole in the cross-correlation with 21 cm Intensity Mapping}

\newcommand{\bk}{{\mathbf k}}

\newcommand{\bx}{{\mathbf x}}

\newcommand{\bn}{{\mathbf n}}

\newcommand{\be}{\begin{equation}}
\newcommand{\ee}{\end{equation}}

\newcommand{\bea}{\begin{eqnarray}}
\newcommand{\eea}{\end{eqnarray}}
\newcommand{\bean}{\begin{eqnarray*}}
\newcommand{\eean}{\end{eqnarray*}}

\newcommand{\HH}{{\cal H}}

\author[a, b]{Francesca~Lepori,}
\author[c, a, b]{Enea~Di~Dio,}
\author[a, b]{Eleonora~Villa,}
\author[a, b, c]{Matteo~Viel}

\affiliation[a]{SISSA- International School for Advanced Studies, \\ Via Bonomea 265, 34136 Trieste, Italy}
\affiliation[b]{INFN, Sezione di Trieste,\\ Via Valerio 2, I-34127 Trieste, Italy}
\affiliation[c]{INAF - Osservatorio Astronomico di Trieste,\\  Via G. B. Tiepolo 11,  I-34143 Trieste, Italy}

\emailAdd{flepori@sissa.it}
\emailAdd{enea.didio@oats.inaf.it}
\emailAdd{evilla@sissa.it}
\emailAdd{viel@sissa.it}

\newcommand{\tj}[6]{ \begin{pmatrix}
   #1 & #2 & #3 \\
   #4 & #5 & #6 
\end{pmatrix}}

\abstract{
We investigate the future perspectives of the detection of the relativistic dipole by cross-correlating the 21 cm emission in
Intensity Mapping (IM) and galaxy surveys at low redshift. We model the neutral hydrogen (HI) and the galaxy population by means of the halo
model to relate the parameters that affect the dipole signal such as the biases of the two tracers and the 
Poissonian noise. 
We investigate the
behavior of the signal-to-noise as a function of the galaxy and magnification biases, for two fixed models of the neutral 
hydrogen. In both cases we found that the signal-to-noise does not grow by increasing the difference between the biases of the two tracers, due to the larger shot-noise yields by highly biased tracers.
We also study and provide an optimal luminosity-threshold galaxy catalogue to enhance the signal-to-noise ratio of the relativistic dipole.
Interestingly, we show that the maximum magnitude provided by the survey does not lead to the maximum signal-to-noise for detecting relativistic effects and we predict the optimal value for the limiting magnitude.  Our work suggests that an optimal analysis could increase the signal-to-noise ratio up to a factor five compared to a standard one.
}

\begin{document}
\maketitle
\flushbottom

\section{Introduction}

The state-of-the-art analysis of the Cosmic Microwave Background (CMB) anisotropies \cite{planck2015} and of the Large Scale Structure (LSS) 
of the universe \cite{Alam:2016hwk} offers robust observational evidence in favor of the standard cosmological model, known as $\Lambda$CDM.  
However, the $\Lambda$CDM model presents some theoretical troubles, above all the cosmological constant problem \cite{ccproblem}
and the nature of the dark components in the matter sector \cite{DM1937, bullett}. These issues keep raising an increasing interest in the scientific community and strongly necessitate further investigation of the foundational principles from which the standard model was built \cite{Bull:2015stt}.
A pivotal role in testing the laws of gravity at cosmological scales will be played by the forthcoming LSS surveys aiming to map with 
stunning precision the large scale structure of the universe. The huge amount of incoming data requires an equally 
powerful advance from the theoretical and modelling side. 
One of the issues related to the correct modelling of the LSS observables is the fact that we do not observe directly the dark matter density field,
but we observe some biased tracers 
at a given observed redshift and direction in the sky.
A fully relativistic computation of what is really measured in a galaxy survey was performed at linear order
\cite{yooA,yooB, bonvin_durrer, challinor_lewis, Schmidt:2012ne, Bertacca:2012, Raccanelli:2013gja, Raccanelli:2013} and at second order in perturbation theory \cite{Yoo:2014sfa, Bertacca:2014dra, Bertacca:2014hwa, DiDio:2014lka, DiDio:2015, Nielsen:2016ldx, Umeh:2016nuh, DiDio:2016kyh, Jolicoeur:2017nyt, Bertacca:2017dzm}. This formalism has been further extended to other observables such as the lensing convergence \cite{Bonvin:2008ni}, the HI brightness temperature \cite{Hall:2012} and the Lyman-$\alpha$ transmitted flux \cite{Irsic:2015}.
Since we measure galaxy positions by collecting photons which have travelled in a clumpy universe, the LSS observables are also affected by other contributions, which were commonly neglected, beyond the standard local overdensity
and Redshift Space Distortion (RSD) \cite{kaiser}.
These terms, which include gravitational lensing convergence, gravitational redshift, Doppler and integrated Sachs-Wolfe effect, may bias the estimation of some cosmological parameters
\cite{Camera:2014bwa, Camera:2014sba, Raccanelli2015, Montanari:2015rga, DiDio:2016ykq, Cardona:2016qxn}, in particular for deviations from $\Lambda$CDM \cite{Villa:2017yfg}.
More interestingly, a measurement of these effects offers the opportunity to extract further
cosmological information, allowing to test the Equivalence Principle or the theory of gravity at the largest observable scales.
Even if all these corrective terms are subdominant on sub-horizon scales with respect to the local density and RSD, it has been shown that it is possible
to isolate some of them, by correlating two different tracers\footnote{It has also been claimed~\cite{Raccanelli:2016} that futuristic surveys may be able to detect them through a single tracer analysis.}~\cite{mcdonald:2009,Yoo:2012se, Bonvin:2013, Alonso:2015sfa,Fonseca:2015laa, Irsic:2015, Bonvin:2015,Gaztanaga:2015, Dai:2015wla, Borzyszkowski:2017ayl, Abramo:2017xnp}.
This can be achieved using the fact that some terms carry an odd symmetry with respect to the line of sight, which can be exploited with different tracers.
In fact, in the same way as RSD introduces an anisotropy that sources other even multipoles
in the correlation function of a LSS tracer apart from the monopole, gravitational redshift and Doppler effects break the symmetry with respect 
to the exchange of two tracers along the line of sight. Therefore, they induce non-vanishing odd multipoles in the cross-correlation function or an imaginary part in the Fourier power spectrum.
Measuring the relativistic effects is a new opportunity for testing the consistency
of general relativity \cite{PhysRevD.87.104019, bonvin2014_isolating} and may offer an alternative method to measure the peculiar velocity 
field of the sources \cite{Hall:2016, Bonvin:2016}. 
The first measurement of the dipole for the cross-correlation of two galaxies populations
\cite{Gaztanaga:2015} was a measurement of the dipole induced by the so-called large-angle effect. 
Nevertheless, at the lowest order in the distant observer approximation, the large-angle effect can be written as a combination of the monopole and the quadrupole and therefore it does not provide new information, at least in the regime where this approximation is reliable.
In this work we will investigate the possibility to detect the relativistic Doppler corrections by cross-correlating galaxies and
21 cm Intensity Mapping (IM). The prospect of measuring the relativistic dipole by cross-correlating these two tracers
was addressed in \cite{Hall:2016}, where a signal-to-noise analysis is presented for specific IM experiments and galaxy survey.
Here we present a survey-independent analysis, with the aim of understanding which types of galaxies are more suitable for
detecting the relativistic dipole and how the signal-to-noise can be optimized by properly choosing the luminosity threshold of 
the galaxy catalogue. Since the Doppler corrections are more relevant at low redshift, this is the regime we are focused on.

Cross-correlation observational studies  between galaxies and IM at low redshift ($z<1$) have already been performed in recent years with the goal of detecting the diffuse neutral hydrogen (HI) \cite{chang10,masui13} and also at high redshift exploiting the cross-correlation between IM and Lyman-break galaxies \cite{crosspaco} and the Lyman-$\alpha$ forest \cite{carucci17} in the post-reionization era. Furthermore, higher redshift investigations of the cross correlations are also of primary importance during or around HI reionization (e.g. \cite{lidz09}).

The paper is structured as follows. In section \ref{Sec:1} we revise the relativistic formalism for galaxy surveys and IM experiments.
Furthermore, we summarize the formalism that will be used in the following sections to compute the multipoles of the cross-correlation for the two tracers and its covariance, and we discuss two possible contaminations to the relativistic dipole: the lensing dipole and the wide-angle effect.
More details on the formalism and the approximation we adopt in this work can be found in the Appendices \ref{Ap_A} and \ref{Ap_B}.
In section \ref{Sec:2} we use the halo model to describe in a fairly general way the HI and galaxy distribution
properties, and we show how the biases of the two tracers can be related to their shot-noise.
In section \ref{Sec:SN} we present a signal-to-noise analysis for the relativistic dipole. In particular, we investigate the behavior of the signal-to-noise
as a function of the galaxy bias and magnification bias, for two HI models. 
In section \ref{Sec:HOD} we model luminosity-threshold galaxy catalogues with an halo occupation distribution model and we studied the
signal-to-noise for the relativistic Doppler dipole as a function of the limiting magnitude.
Finally, in section \ref{Sec:concl} we sum up the results of our work and we draw the conclusions.

We stress that the  rationale of this work is to provide a first quantitative investigation of the optimal strategy to detect relativistic effects by exploiting the 
cross-correlation signal between IM and galaxies, in doing that we will learn that the low redshift regime is important to have a high value of the  signal-to-noise-ratio 
of the relativistic effects. This work is also motivated by the fact that there are indeed wide area low redshift surveys planned or under way that could provide the necessary data for the galaxy populations (e.g. the planned EMU \cite{norris}) or the WISE  data set \cite{bilicki}) to be interfaced with IM data provided by 
radio telescopes like LOFAR \cite{lofar}, Murchison Wide-field Array \cite{mwfa}, GMRT \cite{gmrt}, the Ooty Radio Telescope\cite{gmrt}, CHIME \cite{CHIME}, ASKAP \cite{atnf}, MeerKAT \cite{meerkat} and SKA \cite{ska}. 

Throughout all the paper we assume a spatially flat $\Lambda$CDM cosmology with parameters $h = 0.67556$, $\Omega_\text{cdm} h^2 = 0.12038$, $\Omega_\text{b}h^2 = 0.022032$.
The primordial amplitude and spectral index are $A_\text{s} = 2.215 \times 10^{-9}$ and $n_\text{s} = 0.9619$, respectively.
The matter power spectrum was computed with the Cosmic Linear Anisotropy Solving System ({\sc class}) code~\cite{class1, class2}, with 
pivot scale $k=0.05\,\text{Mpc}^{-1}$.

\section{Cross-correlation odd multipoles}

\label{Sec:1}
In this section we will report the expression for the observable quantities in a galaxy redshift survey and in a 21 cm
intensity mapping experiment and the corresponding cross-correlation.
\looseness=-1 We consider a perturbed Friedmann-Lemaître-Robertson-Walker (FLRW) metric and we work in longitudinal gauge\footnote{Since we consider only observable quantities, the gauge choice will not affect any result.}
\be \label{metric}
ds^2 = a(\eta)^2 \left( -\left( 1+ 2 \Psi \right) d\eta^2 + \left( 1 - 2 \Phi \right) d\bx^2 \right) \, ,
\ee
where $\eta$ denotes the conformal time, $a(\eta)$ is the scale factor, and the metric perturbations, $\Psi$ and $\Phi$, are the Bardeen potentials. 
We also remark that the equations summarized in this section do not assume General Relativity (GR).

\subsection{Galaxy number counts}

In a galaxy clustering experiment we measure the number of galaxies $N \left( \bn , z \right)$ in terms of an angular direction $\bn$,\footnote{$\bn$ denotes here the unit vector (direction) in which the photons propagate, while the angular position in the celestial sphere is $-\bn$.} and a measured redshift $z$. We can then define the galaxy number counts as
\be
\Delta_{\text{gal}} (\mathbf{n}, z) = \frac{N \left( \bn , z \right) - \langle N \rangle \left( z \right) }{\langle N \rangle \left( z \right)} ,
\ee
where $\langle .. \rangle$ denotes the angular average at fixed observed redshift $z$.
The galaxy number counts were computed to first order in perturbation theory ~\cite{yooA,yooB, bonvin_durrer, challinor_lewis}, by accounting that galaxies are a biased tracer of the underline dark matter field,
and they can schematically be expressed\footnote{The expression for the number counts in \eqref{nc_gal} assumes that galaxies follow geodesics
\begin{equation}
\partial_r \Psi = \mathbf{V}'\cdot \mathbf{n} +\HH \mathbf{V}\cdot \mathbf{n} .
\end{equation}
} as~\cite{CLASSgal}
\begin{align}
\Delta_{\text{gal}} (\mathbf{n}, z, m^*) &= b_\text{gal}(z, m^*) D  + \frac{1}{\mathcal{H}(z)} \partial_r (\mathbf{V}\cdot\mathbf{n}) \notag
\label{nc_gal}\\
&  + (5s(m^*, z) - 2) \int_0^{r(z)} \frac{r(z) - r}{2 r(z) r} \Delta_{\Omega} (\Phi + \Psi) dr \notag \\
&+ \Biggl(\frac{\mathcal{H'}}{\mathcal{H}^2} + \frac{2-5 s(m^*, z)}{r\mathcal{H}} + 5 s(m^*, z) - f^{\text{gal}}_{\text{evo}} (m^*, z)
\Biggr)(\mathbf{V}\cdot \mathbf{n} ) \notag \\
&+( f^{\text{gal}}_{\text{evo}}  - 3)\HH V +(5s - 2) \Phi + \Psi + \frac{1}{\mathcal{H}} \Phi' +\frac{2-5s}{r(z)} \int^{r(z)}_0 dr (\Phi + \Psi) \notag \\
& +\Biggl(\frac{\mathcal{H'}}{\mathcal{H}^2} + \frac{2-5 s }{r(z)\mathcal{H}}+5s  - f^{\text{gal}}_{\text{evo}}\Biggr)\Biggl(\Psi + \int^{r(z)}_0 dr (\Phi' + \Psi')\Biggr) \, ,       
\end{align}
\looseness=-1 where $b_\text{gal}(z, m^*)$ is the galaxy bias of the sources
whose magnitudes are smaller than the magnitude limit of the survey $m^*$
\footnote{The magnitude limit of the survey is related to the flux limit through $m^* = -\frac{5}{2} \log_{10} \Bigl[\frac{F^*}{F_0}\Bigr]$,
where $F_0$ is a reference value for the flux.}, assumed to be linear and local; $D$ is the dark matter density fluctuation in synchronous gauge;  $\mathcal{H}=a'/a$ is the conformal Hubble factor;
$\mathbf{V}$ is the peculiar velocity in longitudinal gauge; $V$ the velocity potential defined by ${\bf V} = -{\bf \nabla} V$; $\eta_o$ is the present time and $r(z) = \eta_o -\eta$ is the conformal distance at redshift $z$; $\Delta_{\Omega}$ represents the angular laplacian operator
and a prime denotes a partial derivative with respect to the conformal time.

The bias factors in the expression above, $s(m^*, z)$ and $f_{\text{evo}} (m^*, z)$, are the magnification and the evolution biases
of the galaxy catalogue, respectively. 
The magnification bias $s$ is the slope of the cumulative luminosity function \cite{CLASSgal, Alonso2015_ST}
\begin{equation}
s(m^*, z) = \frac{\partial \log_{10} \bar{N} (z, m< m^*)}{\partial m^*}, \label{s_bias}
\end{equation}
and $\bar{N} (z, m< m^*)$ denotes the cumulative luminosity function
\begin{equation} \label{def:fevo}
\bar{N} (z, m< m^*) = \int^{\infty}_{\ln{L*}} \phi(\eta(z), \ln{L}) d\ln{L},
\end{equation}
where $ \phi(\eta(z), \ln{L})$ is the luminosity function and $L^*$ is the luminosity threshold of the survey that is related to the flux threshold by $L^* = 4 \pi (1+z)^2 r^2(z) F^*$.
The evolution bias describes the departure from a sample of sources conserved in a comoving region 
\begin{equation}
f^{\text{gal}}_{\text{evo}} (m^*, z) = \frac{\partial \ln \bar{N} (z, m< m^*)}{\HH \partial \eta} = -(1+z) \frac{\partial \ln \bar{N} (z, m< m^*)}{\partial z}.
\label{fevo_gal}
\end{equation}
Modelling the magnification and the evolution biases of a galaxy catalogue requires a prior knowledge on the luminosity function 
of the targeted galaxy population. 

\subsection{21 cm brightness temperature fluctuation}

Intensity mapping~\cite{IM_paper} is a novel technique which aims to map the Large Scale Structure of the universe by measuring the collective emission of many galaxies without resolving individual sources. 
21 cm IM experiments target the emission line of neutral atomic hydrogen. The observable quantity is the flux density,
i.e.~the integral of the specific intensity over the solid angle of the telescope beam, which can be related to the HI
brightness temperature in the Rayleigh-Jeans regime (see the Appendix in \cite{Bull2015}  for a more detailed discussion).
The observed fluctuation in the 21 cm brightness temperature has been computed in linear theory in Ref.~\cite{Hall:2012}, including all
the relativistic corrections. The full expression is mathematically equivalent to the expression in \eqref{nc_gal}, with the
magnification bias value set to $s = 2/5$, such that the lensing contribution vanishes.
The fact that the observable in an intensity mapping survey is not affected by gravitational lensing to linear order is due to surface brightness conservation. Indeed the change in the solid angle $d\Omega$ is exactly compensated by the change in the observed flux.
The full expression for the observed fluctuation in the 21 cm brightness temperature is
\begin{align}
\Delta_{\text{21cm}} (\mathbf{n}, z) &= b_\text{HI}(z) D  + \frac{1}{\mathcal{H}(z)} \partial_r (\mathbf{V}\cdot\mathbf{n}) 
+ \Biggl(\frac{\mathcal{H'}}{\mathcal{H}^2} + 2 - f^{\text{HI}}_{\text{evo}} (z)
\Biggr)(\mathbf{V}\cdot \mathbf{n} )\label{T_21} +( f^{\text{HI}}_{\text{evo}}  - 3)\HH V  \notag \\
& + \Psi + \frac{1}{\mathcal{H}} \Phi'  +\Biggl(\frac{\mathcal{H'}}{\mathcal{H}^2} + 2  
- f^{\text{HI}}_{\text{evo}}\Biggr)\Biggl(\Psi + \int^{r(z)}_0 dr (\Phi' + \Psi')\Biggr), 
\end{align}
where $b_\text{HI}(z)$ and $f^{\text{HI}}_{\text{evo}}(z)$ are the bias and the evolution bias of the neutral hydrogen, respectively.

The evolution bias for the HI can be defined similarly to the galaxy evolution bias in \eqref{def:fevo}. 
Taking into account that we observe all the HI emissions from a patch of the sky, it depends on the redshift evolution
of the HI comoving density $\bar \rho_\text{HI}$ (see the Appendix in \cite{Alonso2015_ST})
\begin{equation}
f^{\text{HI}}_{\text{evo}} =  \frac{\partial \ln \bar \rho_{\text{HI}}( z)}{\HH \partial \eta} = -(1+ z) \frac{\partial \ln \bar\rho_{\text{HI}}( z)}{\partial  z}.
\label{fevo_HI}
\end{equation}

\subsection{21 cm - Galaxies cross-correlation}
\label{sub2.3}
Cross-correlation studies are promising techniques to study relativistic effects.
In fact, in the past years it has been pointed out that relativistic effects source odd multiples of the correlation function
or Fourier space power spectrum, when two different tracers are cross-correlated~\cite{mcdonald:2009,Yoo:2012se, Bonvin:2013,Alonso:2015sfa,Fonseca:2015laa, Irsic:2015, Bonvin:2015,Gaztanaga:2015,Borzyszkowski:2017ayl}.

\begin{figure}
\begin{center}
\begin{tikzpicture}[] 
\node[draw,circle] (B) at (0,-2.5) {\textsc{Obs}};
\node[] (A1) at (1.7,-2.4) {$-\mathbf{n_1}$};
\node[] (A2) at (0.8,-1.0) {$-\mathbf{n_2}$};
\node[] (A2) at (6.5, 4.7) {$z_2$};
\node[] (A2) at (6.5, -1.6) {$z_1$};
\node[] (A) at (1.7,-1.6) {$-\mathbf{n}$};
\draw[->, thick]  (B) -- (1.8, -0.55);
\draw[->, thick]  (B) -- (2.5, -1.875);
\node[draw,circle] (gal) at (6, 4.0) {{\textsc{Gal}}};
\node[] (zm) at (6, 1.5) {};
\fill (zm) circle [radius=2pt];
\node[] (test) at (0.0, 5.0){};
\node[] (beta) at (8.0,2.83333333333) {};
\draw[dashed, -, thick]  (B) -- (beta);
\draw[dashed, -, thick] (B) -- (zm) node [midway, above, sloped] (TextNode) {$r(z_\text{m})$};;
\node[draw,circle] (h1) at (6, -1.0) {{\textsc{HI}}};
\draw[dashed, -, thick]  (B) -- (gal);
\draw[dashed, -, thick] (B) -- (gal) node [midway, above, sloped] (TextNode) {$r_2$};;
\draw[dashed, -, thick]  (B) -- (h1);
\draw[dashed, -, thick] (B) -- (h1) node [midway, below, sloped] (TextNode) {$r_1$};;
\draw[dashed, -, thick]  (h1) -- (gal);
\draw[->, thick]  (B) -- (2.2, -1.03333333333);
\draw[->, thick]  (h1) -- (6, 0.9);
\draw[<->, thick]  (7, -1.0) -- (7, 4.);
\node[] (dist) at (7.5,1.5) {d};
\node[] (N) at (6.3, 0.0) {$\mathbf{N}$};
\draw[->, thick]  (B) -- (test);
\draw[thick] ([shift=(33:1cm)]6,1.5) arc (33:90:1cm);
\node[] (b) at (6.6,2.7) {$\beta$};
\draw[thick] ([shift=(14:2.9cm)]0,-2.5) arc (14:47:2.9cm);
\draw[thick] ([shift=(14:3cm)]0,-2.5) arc (14:47:3cm);
\node[] (X) at (3.0,-1.1) {$\theta$};
\end{tikzpicture}
\end{center} 
 \caption {Illustration of the position of the two tracers under investigation with respect to the observer, in terms of the observed coordinates $\mathbf{z}_1$, $\mathbf{z}_2$ and $\theta$ and the coordinate system adopted in this work.
}
\label{scheme}
\end{figure}
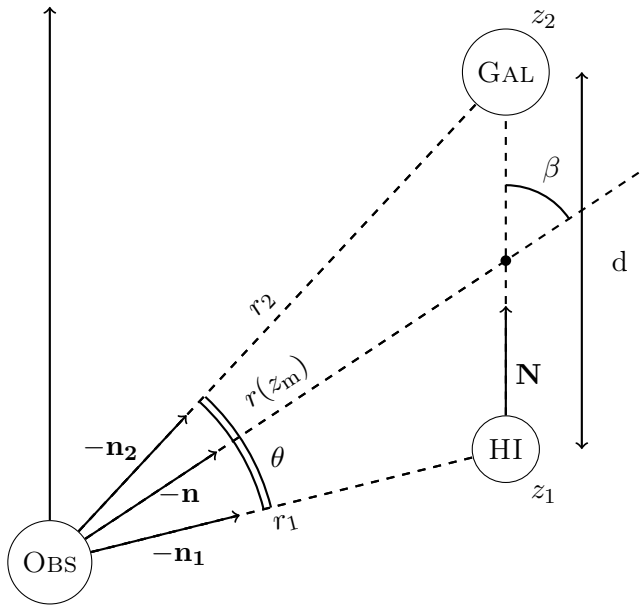

The two-point cross-correlation between the HI temperature and galaxy number count fluctuations, in terms of the observed coordinates, is defined as
\begin{equation}
\xi^{\text{HI}, \text{gal}}(z_1, z_2, \theta) =  \mean{\Delta_\text{21cm}(\mathbf{n}_1, z_1)\Delta_\text{gal}(\mathbf{n}_2, z_2)}, \qquad \cos{\theta} \equiv \mathbf{n_1}
\cdot \mathbf{n_2}, \label{full_cross}
\end{equation}
where the $\mean{...}$ denotes the ensemble average, replaced in observation by the average over observed directions at a fixed observed redshift.
In figure~\ref{scheme} we represent a scheme of the observed coordinates for the system under investigation.

The observed coordinates $z_1, z_2$ and $\theta$ can be converted, by assuming a cosmology, into comoving distances.
We denote with $r_1$ and $r_2$ the comoving distance at $z_1$ and $z_2$, respectively, 
while $-\bn_1$ and $-\bn_2$ are the unit vectors pointing in the direction of the two tracers (HI and galaxies, respectively). 
Due to angular statistical isotropy, the cross-correlation can be the written in terms of three coordinates. In this work, we will adopt the following coordinate system
(see figure \ref{scheme} for a schematic representation): the comoving separation $r$ at the mean redshift $z_m = (z_1 + z_2 ) /2$, the separation between the two sources $d$ and the angle $\beta$ (or $\mu = \cos \beta$), where $\cos \beta = -\bn \cdot \mathbf{N}$, while $-\bn$ and 
$\mathbf{N}$ are the unit vectors pointing in the direction of the mean redshift and the distance between the two tracers, respectively, as defined in figure~\ref{scheme}.

The cross-correlation function can be then expanded into Legendre polynomials $L_{\ell}$, 
\begin{equation}
\xi^{\text{HI}, \text{gal}}(r, d, \mu) = \sum_{\ell} \xi_{\ell}(r, d) L_{\ell}(\mu),
\end{equation}
and the coefficients of this expansion, the multipoles of the correlation function, are defined as
\begin{equation}
 \xi_{\ell}(r, d) = \frac{2\ell + 1}{2} \int_{-1}^{1} \xi^{\text{HI}, \text{gal}}(r, d, \mu) L_{\ell} (\mu) d\mu.  \label{eq213}
\end{equation}
This definition does not include an optimal weight based on the galaxy number density, 
that is generally included in order to reduce the shot-noise \cite{Ross:2012qm}. We do not include this effect for the following reasons. First, the weight is survey dependent, therefore a correct modelling of this effect will require to make some assumptions
on the survey specifics. Furthermore, this weight is expected to reduce the noise, therefore our analysis without any assumption on the
weight can be considered conservative.

In a similar way we can define\footnote{We neglect the redshift evolution in the following definition. As shown in Ref.~\cite{Bonvin:2013} the redshift evolution corrections are subdominant compared to the wide-angle dipole contamination of the standard terms.
We also do not consider the integrated terms here. While time-delay and ISW effects are negligible, lensing magnification may contaminate the measurement of the Doppler dipole. We therefore study the contamination of magnification lensing in Sec.~\ref{LensWA}.}
 the Fourier cross power spectrum $P^{\text{HI}, \text{gal}}(z_\text{m}, \bk)$ as 
\begin{equation}
\mean{\Delta_\text{21cm}(\bk_1, z_1)\Delta_\text{gal}(\bk_2, z_2)} = (2\pi)^3 P^{\text{HI}, \text{gal}}(z_\text{m}, \bk_1) \delta_D(\bk_1+ \bk_2), 
\label{full_ps}
\end{equation}
where $\Delta_\text{21cm}(\bk_1, z_1)$ and $\Delta_\text{gal}(\bk_2, z_2)$ denotes the Fourier transform of
$\Delta_\text{21cm}(\mathbf{n}_1, z_1)$ and $\Delta_\text{gal}(\mathbf{n}_2, z_2)$, respectively, and
$\delta_D$ is the Dirac delta.
The multipoles of the power spectrum can be computed similarly to the correlation function multipoles, and they are 
proportional to the linear matter power spectrum at the mean redshift, $P(k, z_\text{m})$,
\begin{equation} \label{matterpower}
\mean{D(\bk_1, z_\text{m})D(\bk_2, z_\text{m})} = (2\pi)^3 P(k_1, z_\text{m}) \delta_D(\bk_1+ \bk_2).
\end{equation}
In the distant observer limit, i.e.~$d\ll r$, the angular position $-\bn$ is assumed to be fixed for the two observed sources
$\bn_1 = \bn_2 = \bn$ and the full-sky
correlation function can be simplified. Indeed, the full expression of the correlation function (Eq.~\eqref{full_sky_cf} in Appendix \ref{Ap_A}) can be written as power series expansion in $d/r$, and 
by taking only the lowest order (i.e.~assuming $d \ll r$) the multipoles of the correlation function can be expressed in terms of multipoles of the power spectrum \cite{Hall:2016},
\begin{equation} 
\xi_{\ell}(d, r(z_\text{m})) = (-i)^{\ell} \int \frac{k^2 dk}{2\pi^2} P_{\ell}(k, z_\text{m}) j_{\ell} (k\,d), \label{multipoles}
\end{equation}
where
\be
P_{\ell}(k,z_\text{m}) =  \frac{2\ell + 1}{2} \int_{-1}^{1} P^{ {\text{HI}, \text{gal}}}(k, z_\text{m}, \mu) L_{\ell} (\mu)d \mu
\ee
are the coefficients of the expansion of the angle dependent power spectrum in Legendre polynomials, at fixed redshift 
$z$, and $j_{\ell}$ are the spherical Bessel function of order $\ell$.
Following the same strategy adopted in \cite{Hall:2016}, we consider a local expansion of the multipoles of the power spectrum in
power of ($\HH/k$) and we include the leading terms with respect to the expansion parameter ($\HH/k$).
The even multipoles are dominated by the Newtonian contribution (i.e.~the first line in \eqref{nc_gal} and the first two terms in \eqref{T_21}),
which are simply proportional to the matter power spectrum,
whilst the odd multipoles are suppressed by a factor ($\HH/k$), which is provided by the correlation of Doppler contribution with density and redshift space distortions (i.e.~the third line in Eq.~\eqref{nc_gal} and the third term in Eq.~\eqref{T_21}). 
For the cross-correlation of galaxies and 21 cm brightness temperature, the leading contributions to the power spectrum multipoles are
\begin{align}
P_0 (k) = &\Biggl[b_{\text{HI}} b_{gal} + \frac{f}{3}(b_{\text{HI}} + b_\text{gal} ) + \frac{f^2}{5}\Biggr] P(k) \, , \notag \label{Pell} \\  
P_1 (k) = &\, (- i) \, \Biggl[\Bigl(b_\text{gal}  C_\text{HI}  - b_\text{HI} C_\text{gal} \Bigr) f + \frac{3}{5} \Bigl(C_\text{HI}-C_\text{gal} \Bigr) f^2
\Biggr] \frac{\HH}{k} P(k) \, , \notag \\
P_2 (k) =  &\Biggl[ \frac{2}{3}f (b_{\text{HI}} + b_\text{gal} ) + \frac{4}{7} f^2 \Biggr] P(k)\, , \notag \\
P_3 (k) = &\, i \, \frac{2}{5}\Bigl(C_\text{gal} - C_\text{HI}\Bigr) f^2 \frac{\HH}{k} P(k)\, , \notag \\
P_4 (k) = & \frac{8}{35} f^2 P(k), 
\end{align} 
where $f= d\ln D /d\ln a$ is the growth factor. In $\Lambda$CDM the growth factor is given by $f(z) = \Omega_m(z)^{4/7}$ \cite{Lahav91, Carrol92}.
In Eq.~\eqref{Pell} the mean redshift is assumed to be fixed and, for the sake of simplicity, it is omitted from the notation.
The coefficients $C_\text{gal}$ and $C_\text{HI}$ are defined as
\begin{align} \notag
C_\text{gal} &= \Biggl( \frac{\mathcal{H'}}{\mathcal{H}^2} + \frac{2-5 s}{r\mathcal{H}} + 5 s - f^{\text{gal}}_{\text{evo}} \Biggr), \label{coeff} \\
C_\text{HI}  &=   \Biggl(\frac{\mathcal{H'}}{\mathcal{H}^2} + 2 - f^{\text{HI}}_{\text{evo}} (z) \Biggr). 
\end{align}

\begin{figure}
\includegraphics[width=\textwidth]{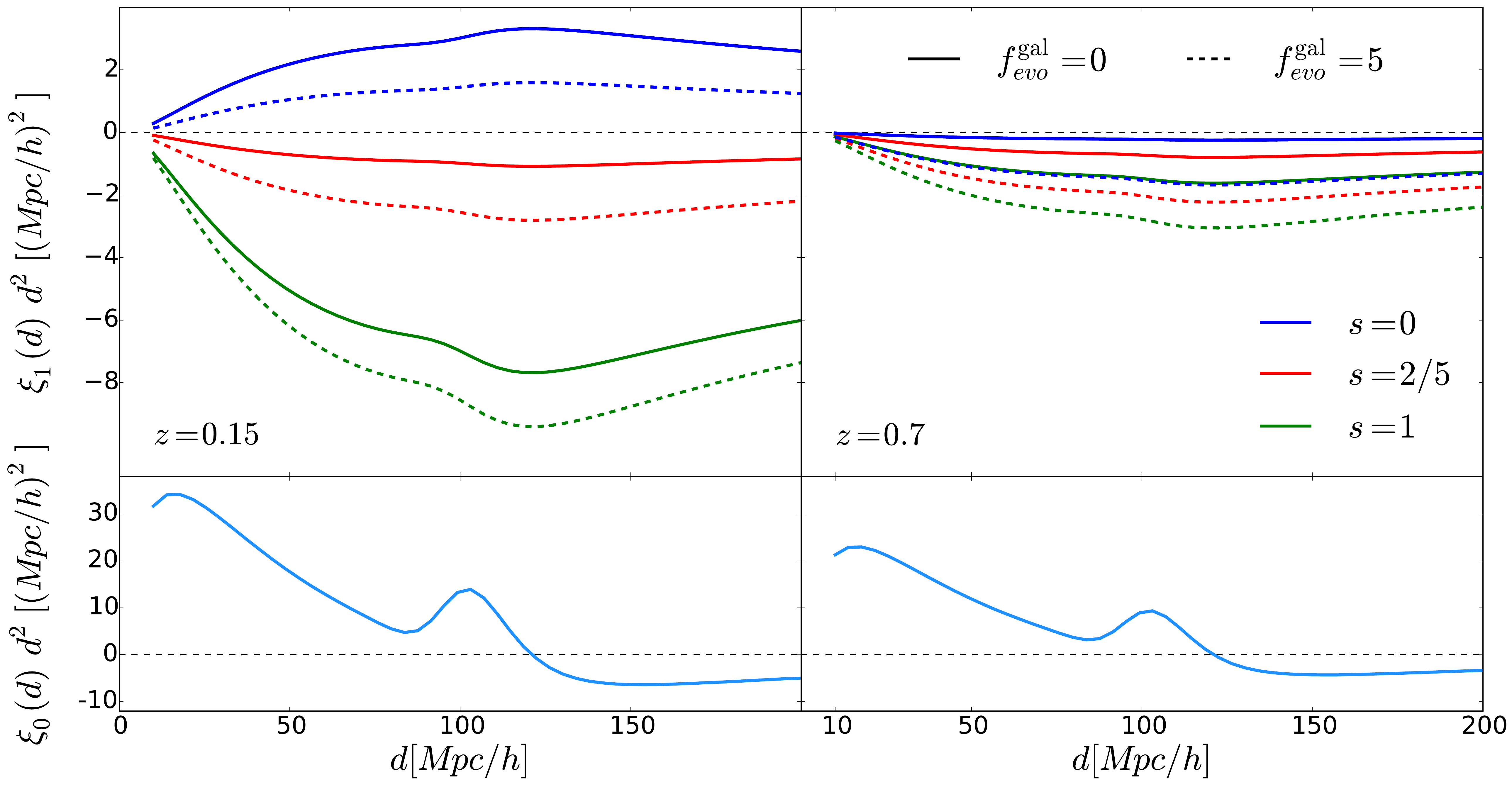}   
    \caption{Cross-correlation dipole (top panels) and monopole (bottom panels), computed from Eq.~\eqref{multipoles}, 
    at $z = 0.15$ (left panel) and $z = 0.7$ (right panel). In the top panels different colors denotes different values for the magnification bias of the galaxy catalogue, whilst different line-styles refer to two different values of the galaxy evolution bias. The clustering biases are set to the values $b_\text{gal} = 1$ and $b_\text{HI} = 0.6$, the evolution bias for the HI is set to be $f^{\text{HI}}_{\text{evo}}  = -1.5$.
    }
\label{fig1}
\end{figure}

In figure \ref{fig1} we compare the cross-correlation dipole (top panels) and monopole (bottom panels)
at two different mean redshifts $z_\text{m} = 0.15$ (left panel) and $z_\text{m} = 0.7$ (right panel). 
In this plot, we set the values of the clustering biases to be $b_\text{gal} = 1$ and $b_\text{HI} = 0.6$.
The evolution bias for the HI is assumed to be $f^{\text{HI}}_{\text{evo}}  = -1.5$, while we denote with
different colors different values of the magnification bias $s$ and with different line-styles two different values of the galaxy evolution bias. Interestingly, we remark that the sign of the dipole depends strongly on the magnification bias factor and, therefore, it can not be omitted in the analysis.

We see that both the dipole and the monopole signals decrease at larger redshift and that the monopole 
is significantly larger, in amplitude, than the dipole. 
Furthermore, at $z_\text{m} = 0.7$ the terms depending on the galaxy evolution bias become dominant in the dipole. 
Since modelling the evolution properties of a galaxy population is not
an easy task, the cosmological information we can extrapolate from the dipole at large redshift can be contaminated and limited from a prior knowledge about the evolution bias.
The decrease of the dipole at larger redshift depends on two elements: the time evolution of the linear matter power spectrum 
(which affects the monopole as well) and the terms
in the coefficients \eqref{coeff} proportional to $\left( r\mathcal{H}\right)^{-1}$. Therefore, in order to detect relativistic effects we expect an ideal galaxy catalogue in the low redshift regime, and for this reason we will set the mean redshift of observation for our analysis to be $z_\text{m} = 0.15$.

\subsection{Contaminations to the relativistic dipole}
\label{LensWA}
The dipole of the cross-correlation that we discussed in the previous section is sourced by the Doppler corrections to the galaxies number counts and to the
observed brightness temperature of the 21 cm emission. This is usually considered the main contribution to the dipole of the cross-correlation between two tracers,
in fact the Doppler corrections depend on the projection of the peculiar velocity along the line of sight and therefore they are intrinsically anisotropic.
 
Nevertheless, a measurement of the dipole would be contaminated by other sources of anisotropy~\cite{Bonvin:2013}.
In this section we will discuss two possible contaminations: the dipole induced by gravitational lensing and the wide-angle effects. 
The latter have been extensively studied in Refs.~\cite{Szalay:1997cc, Szapudi:2004gh, Papai:2008bd, Matsubara:1999, Bharadwaj:1998bq, Raccanelli:2010hk, Samushia:2011cs, Bertacca:2012,Raccanelli:2013, Raccanelli:2012gt}.
Beside these two contaminants, there are further corrections induced by the redshift evolution of the bias and the growth factors. 
They are generally subdominant with respect to the wide-angle correction \cite{Bonvin:2013}, therefore they will be neglected.

The gravitational lensing asymmetry \cite{Bonvin:2013, Bonvin:2016} comes from the cross-correlation of the HI density and the gravitational lensing term in the galaxy number counts
\begin{equation}
\xi^\text{lens}(r, d, \beta) \equiv \mean{ \left( b_\text{HI}(z_1) D \left( \bn_1, z_1 \right)\right)  \left( \frac{5s - 2}{2} \int_0^{r_2} \frac{r_2 - r'}{ r_2 r'} \Delta_{\Omega} (\Phi + \Psi)\left( \bn_2, z' \right) dr'  \right)}, \label{lens}
\end{equation}
and it emerges from the fact that galaxies behind an HI overdensity with respect to the observer will be lensed, while the HI temperature fluctuations are not lensed, to linear order, by the galaxies in front.
The lensing correlation function defined above has been computed in \cite{Bonvin:2013} and further studied in \cite{Bonvin:2016}. 
In the Limber approximation and to the lower order in $d/r$, it reads 
\begin{equation}
\xi^\text{lens}(r, d, \beta) = (1+z_\text{m}) \frac{3\Omega_m \pi}{4}
b_\text{HI} (5s - 2) \, d \, \mathcal{H}_0\, \cos{(\beta)} \,\Theta(r_2-r_1) \mu_\text{lens}(\beta), \label{lens_xi}
\end{equation}
where $\Theta$ is the Heaviside function, $D_1$ is the linear growth factor and the function $\mu_\text{lens}$ is
\begin{equation}
\mu_\text{lens}(\beta) =  \int_{0}^{\infty} \frac{k_\perp dk_\perp}{2\pi^2}\HH_0 P(k_\perp) J_0(k_\perp\, d \,\sin{(\beta)}),
\end{equation}
 being $J_0$ is the order-0 Bessel function. 
The lensing dipole can therefore be computed similarly to the relativistic dipole 
\begin{equation}
\xi^\text{lens}_{1}(r, d) = \frac{3}{2} \int^{1}_{-1} \xi^\text{lens}(r, d,  \mu) L_1(\mu) d\mu.
\end{equation} 
The lensing dipole is proportional to the radial distance between the two tracers and the redshift. Therefore, we expect it to be relevant
in the high redshift regime, in particular because the Doppler dipole decreases with redshift.
\begin{figure}[t]
\begin{center}
\includegraphics[width=0.65\textwidth]{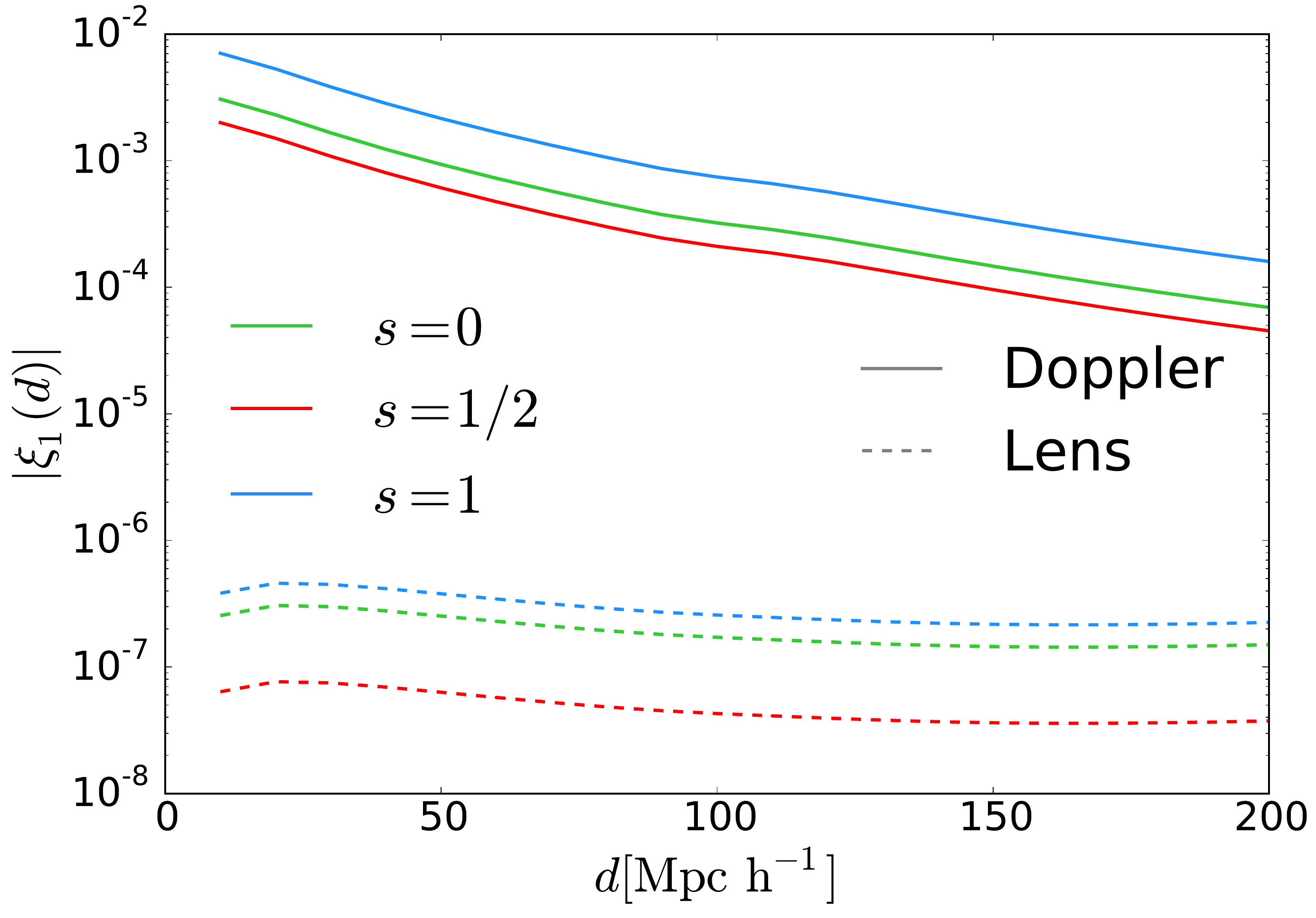}
\caption{Comparison between the amplitude of lensing dipole (dashed lines) and the Doppler dipole (continuous lines) at $z_\text{m} = 0.15$.
The clustering bias of the two tracers are fixed to the values $b_\text{HI} = 0.6$ and $b_\text{gal} = 1$. 
Different colors denote different values of the magnification bias. The HI evolution bias is  $f^\text{evo}_\text{HI} = -1.5$, 
while we fix the galaxy evolution bias to be zero.}
\label{Fig_lens}
\end{center}
\end{figure}
In figure \ref{Fig_lens} we show the amplitude of the lensing dipole compared to the dipole sourced by the Doppler terms,
at redshift $z_\text{m} = 0.15$.
We see that the lensing dipole is always few orders of magnitude smaller then the Doppler dipole, in this redshift regime.
Therefore, in the rest of the paper it will be neglected.

The second correction we will discuss here is the wide-angle effect. In the limit $d\ll r$, we can expand the full-sky correlation function in power of $d/r$. Therefore the Taylor expansion of all the functions of $r_1$ and $r_2$ around $r$ in the leading terms, namely density and redshift space distortions, induces a non-vanishing dipole suppressed by $d/r$. Considering that the largest contributions to the relativistic dipole at low redshift is of the order $d/r \sim 1/\left( k r \right)\sim \HH/k$, we need to account for the wide-angle contamination in the dipole.
The leak from the monopole to the dipole due to the wide-angle contribution, at fixed
redshift and at the lowest order in $d/r$, is described by \cite{Bonvin:2013, Hall:2016}
\begin{equation}
\xi^\text{WA}_1(d) = \frac{2f}{5}(b_\text{gal} - b_\text{HI})\frac{d}{r} \int \frac{k^2 dk}{2\pi^2} \label{wa0}
P(k) j_2(k\,d).
\end{equation}
This correction can be written as a combination of the
quadrupole of the autocorrelation of two tracers. Therefore, we correct the dipole estimator $\hat \xi_1$ for
the bias due to the wide-angle effect \cite{Hall:2016} 
\begin{equation}
\hat\xi_1(d, r) \rightarrow \hat\xi_1(d, r) - \frac{3}{10}(\hat\xi^{gal}_2 - \hat\xi^\text{HI}_2)\frac{d}{r}. \label{wa_corr}
\end{equation}

\begin{figure}[t]
\begin{center}
\includegraphics[width=0.9\textwidth]{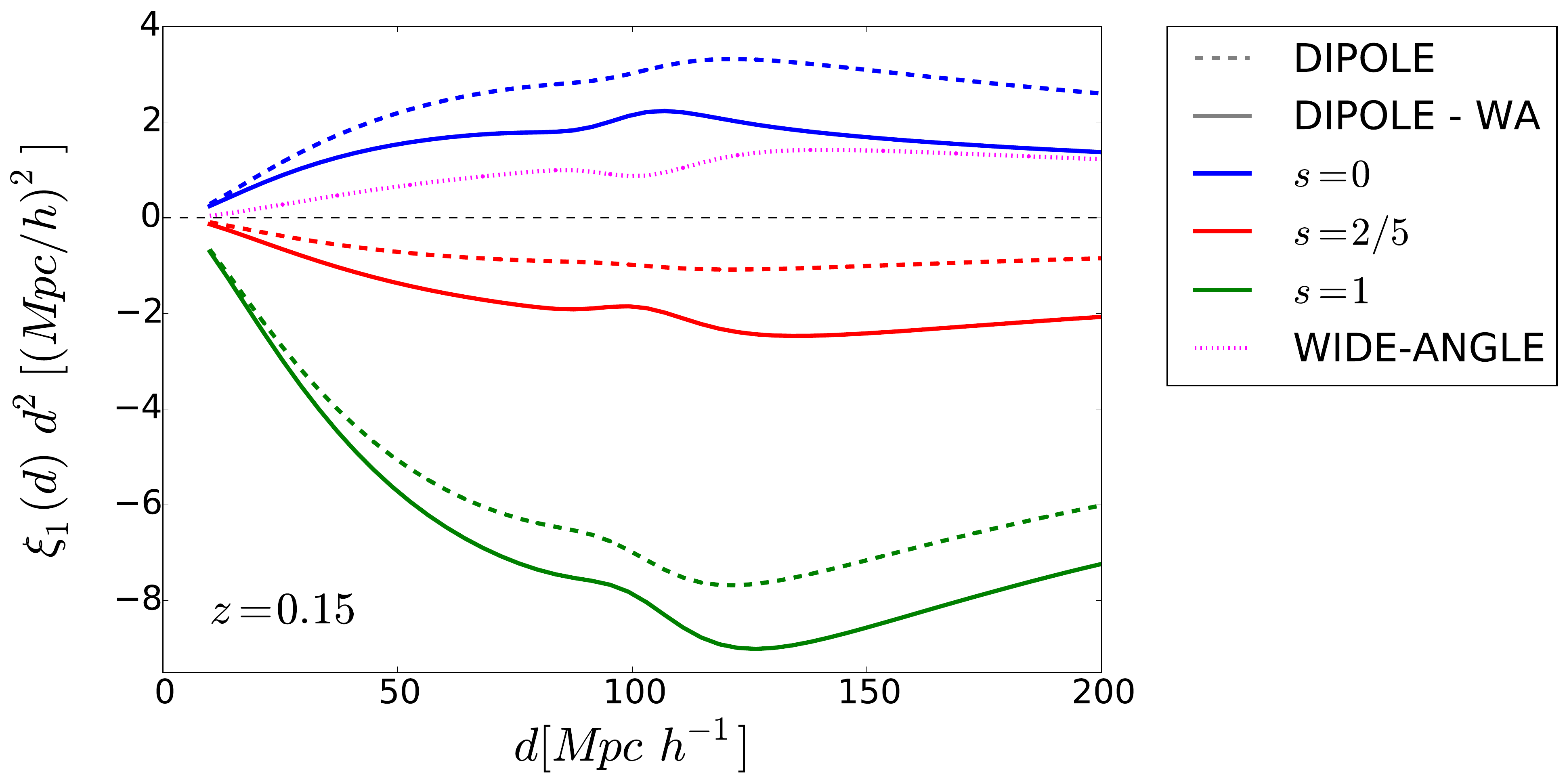}
\caption{Cross-correlation dipole with (continuous line) and without (dashed line) the wide-angle correction in \eqref{wa_corr}.
Different colors denotes different values of the galaxy magnification bias. The clustering biases are set to be 
$b_\text{gal} = 1$ and $b_\text{HI} = 0.6$. The evolution bias is set $f^\text{evo}_\text{HI} = -1.5$ for the neutral hydrogen, while for galaxies it is set to be zero. The magenta dotted line represents the wide-angle correction in \eqref{wa0}.
}
\label{Fig_WA}
\end{center}
\end{figure}
In figure \ref{Fig_WA} we show how the signal changes when the wide-angle correction in Eq.~\eqref{wa_corr} is applied to the estimator.
The magenta dotted line represent the wide-angle contribution to the dipole, computed from Eq.~\eqref{wa0}. Its magnitude
is comparable to the one of the relativistic dipole, thus it is clearly a not negligible contribution.
Furthermore we see that correcting for the wide-angle effect does not necessarily reduces the signal: if its sign agrees with
the one of the dipole, the signal is boosted.
This leads also to an extra contribution to the covariance of the estimator \cite{Hall:2016}.
In the next section we will show how to compute the full covariance of the estimator, the explicit contribution
to the covariance of the wide-angle correction can be found in Appendix \ref{Ap_B}.

\subsection{Covariance for the cross-correlation dipole}

The full covariance matrix for the multipoles of the 2-point correlation function (2PCF) is presented in \cite{Grieb:2015bia} for the single tracer
case and in \cite{Hall:2016} for the multiple tracers case.
Here we will apply the generic expression in \cite{Hall:2016} (Eq.~17) to the dipole.

\begin{align}
\mbox{COV}(d_1, d_2) = &-\frac{9}{V}\int \frac{k^2 dk}{2\pi^2} j_1(k\,d_1) j_1(k\,d_2) \Biggl(\frac{1}{5} P^2_1(k) + \frac{8}{35} 
P_1(k) P_3(k) + \frac{23}{315} P^2_3(k)\Biggr) \notag \\
\notag
+ &  \frac{9}{V}\int \frac{k^2 dk}{2\pi^2} j_1(k\,d_1) j_1(k\,d_2) \Biggl[\frac{1}{3} \Biggl( N_\text{HI} P^\text{gal}_0(k) +
 \frac{1}{n_\text{gal}} P^\text{HI}_0(k) \Biggr)  \\
 \notag 
 + & \frac{2}{15} \Biggl(N_\text{HI} P^\text{gal}_2(k) +
 \frac{1}{n_\text{gal}} P^\text{HI}_2(k) \Biggr)\Biggr] \\
 \notag
 + &\delta_{d_1, d_2 } \frac{3}{4\pi V d_1 d_2 L_p} N_\text{HI} \frac{1}{n_\text{gal}} \\
 \notag
 + & \frac{9}{V} P_\text{N} \int \frac{k^2 dk}{2\pi^2} j_1(k\,d_1) j_1(k\,d_2)  \Biggl(\frac{1}{3} P^\text{gal}_0(k) + \frac{2}{15} P^\text{gal}_2(k)
 \Biggr)\\
 + &\delta_{d_1, d_2 } \frac{3}{4\pi V d_1 d_2 L_p} P_\text{N} \frac{1}{n_\text{gal}}, \label{COV}
\end{align}
where $V$ is the overlapping volume of the galaxy and the intensity mapping surveys;
$P^\text{gal}_\ell$ and $P^\text{HI}_\ell$ denotes the galaxy and the HI power spectrum multipoles, respectively;  $n_\text{gal}$ 
is the comoving number density of galaxies;
$N_\text{HI}$ and $P_\text{N}$ are the shot-noise and the interferometer noise for the HI, respectively; $L_{\rm p}$ denotes the resolution of the IM survey and $\delta_{x, y}$ is the Kronecker delta.

The first line represents the purely cosmic variance contribution to the covariance, the second and the third lines are the cosmic variance - Poisson noise terms, the fourth line is the purely Poissonian contribution, 
while the last two lines are the interferometer noise - cosmic variance term and the interferometer noise - galaxy Poisson noise term, 
respectively.
The terms, which do not involve the integral of the power spectrum, have been integrated by using the orthogonality relation of the spherical Bessel functions
\begin{equation}
\int^{\infty}_0 dk k^2 j_1(k d_1) j_1(k d_2) =  \frac{\pi}{2 d_1 d_2} \delta_D(d_1 - d_2),
\end{equation}
where in the discrete limit \cite{Hall:2016} $ \delta_D(d_1 - d_2) \rightarrow \delta_{d_1, d_2}/L_p$.
The other integrals have been solved numerically, and a smooth cutoff is applied in all the integrals to model the finite resolution of the 
interferometer. To be more precise, all the integrands are multiplied by a top-hat filter in Fourier space $W^4 \! \left( k R\right)$, defined as
\begin{equation}
W(kR) = \frac{3[\sin{(kR)} - kR\cos{(kR)}]}{(kR)^3} , \label{tophat_filter}
\end{equation}
where the scale $R$ is set to be the size of the pixel for the IM, $L_\text{p}$. 

\begin{figure}[t]
\begin{center}
\includegraphics[width=\textwidth]{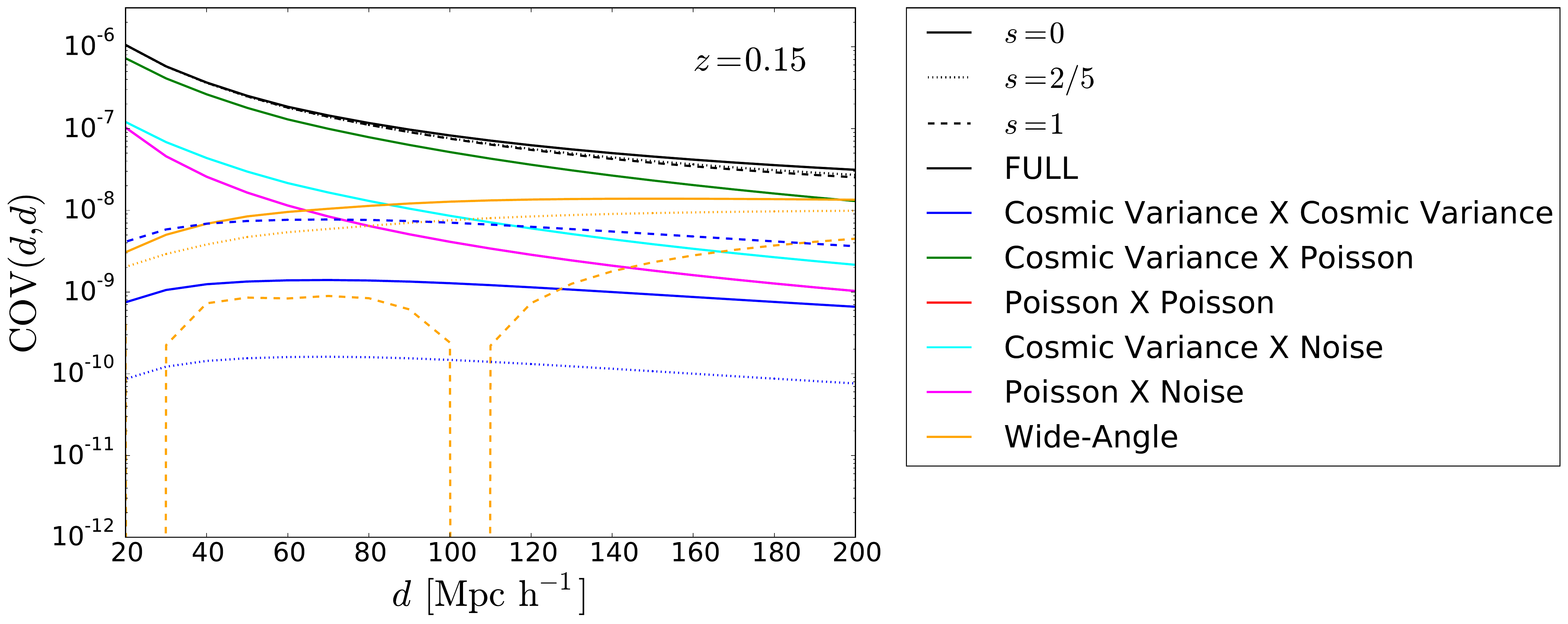}
\caption{
Diagonal entries of the covariance, computed from \eqref{COV}. Different colors denote different terms contributing to the covariance.
The full covariance is the black lines. Different line-styles represent different values of the magnification bias, which affects only the
cosmic variance X cosmic variance contribution (blue line). 
}
\label{cov_diag}
\end{center}
\end{figure}

Figure \ref{cov_diag} shows the different terms contributing to the diagonal covariance entries, at redshift $z_\text{m}=0.15$.
The bias values for the two tracers are the same adopted in figure~\ref{Fig_WA}. Furthermore, we set the shot-noise and the interferometer noise for the HI to $N_\text{HI} = 100\, (\text{Mpc}/\text{h})^3$ and $P_\text{N} = 100 \, (\text{Mpc}/\text{h})^3$, respectively.
The comoving galaxy number density is assumed to be $n_\text{gal} = 10^{-3}  \,(\text{h}/\text{Mpc})^3$ and the volume of the survey is computed 
assuming that the sky coverage of the cross-correlation is $f_\text{sky} = 0.2$ and that the redshift bin is $z \in [0.05, 0.25]$.
The size of the pixel is chosen to be $L_\text{p} = 2\, \text{Mpc}/\text{h}$.
At small scales, the dominant components of the covariance are the terms in which the cosmic variance is cross-correlated with the noise.
On large scales, the covariance introduced by the wide-angle correction becomes more important and on scales $\sim 190 \, \text{Mpc}/{h}$
or larger is the dominant contributor (see the Appendix \ref{Ap_B} for details on how to estimate the covariance introduced by the wide-angle correction). 
The magnification bias enters only in the computation of the purely cosmic variance term, but since this term result to be subdominant 
it does not affect significantly the full covariance (at least its diagonal components).

Both the interferometer noise $P_\text{N}$ and the size of the pixel $L_\text{p} $ depend on the specifics of the IM survey (see \cite{Bull2015, Hall:2016}). 
In order to stay as general as possible, we choose fiducial values ($P_\text{N} = 100 \, (\text{Mpc}/{h})^3$ and $L_\text{p} = 2 \,\text{Mpc}/{h}$, respectively).
They are approximately the values of a survey similar to the
Canadian Hydrogen Intensity Mapping Experiment (CHIME) \cite{CHIME, Bandura:2014gwa}, extrapolated to low redshifts ($z \sim 0.1-0.2$). 

\section{Halo Model approach}
\label{Sec:2}
As shown in section \ref{Sec:1}, the signal and the covariance for the dipole depend on many different parameters 
(clustering and evolution biases, galaxy magnification bias, shot-noise of the two tracers, sky fraction of the cross-correlation and redshift range).
In this section we aim to model the properties of the two tracers in order to find a relation between these parameters, which
are not all independent.
This will allow us to find the optimal specification to look for relativistic effects detection in LSS. Moreover, by determining the relation between all these different parameters we can parameterize better their uncertainties if we need to marginalize over them.
In order to model a generic galaxy population and the neutral hydrogen distribution we will adopt the so-called halo model 
(see \cite{cooray:2002} for a general review).
The halo model was first proposed for modelling the galaxies properties \cite{Seljak:2000, Peacock:2000, Scoccimarro:2000}, but
it has been more recently successfully applied to neutral hydrogen \cite{Villaescusa-Navarro:2014, Castorina:2016, Padmanabhan:2016, Padmanabhan:2017}.
With this approach we are able to model the relation between the tracers clustering bias and their shot-noises.

\subsection{HI model}
\label{HI_sec}
We adopt for the neutral hydrogen the model based on \cite{Castorina:2016}.
As supported by numerical simulations 
\cite{Villaescusa-Navarro:2014}, we can safely assume that the contribution of neutral hydrogen outside the dark matter halos is negligible. Hence, the neutral hydrogen comoving density $\bar \rho_\text{HI}$ at a given redshift $z$ can be computed as
\begin{equation}
\label{OmHI}
\bar\rho_\text{HI} (z) = \int^{\infty}_{0} n(M, z) M_\text{HI}(M,z) dM,
\end{equation}
where $n(M,z)$ is the halo mass function at redshift $z$, i.e.~the comoving number density of halos with masses in the range
between $M$ and $M + dM$, $M_\text{HI}(M,z)$ is the average HI mass in a halo of
mass $M$ at redshift $z$. 
The halo mass function can be expressed as
\begin{equation}
n(M, z) = - \frac{\bar \rho_m}{M^2}f(\sigma)\frac{d\ln{\sigma}}{d\ln{M}}, 
\end{equation}
where $\bar\rho_m = \Omega_m (z)\,\rho_c^0$, $\rho_c^0$ is the critical density at $z = 0$, 
$\sigma$ is the root mean square of the variance of the linear density field, smoothed on the scale $R(M)$, the radius enclosing an amount of mass equal to $M$
\begin{equation}
\sigma^2(R, z) = \frac{1}{2\pi^2} \int^{\infty}_0 k^2 P(k,z) W^2(kR) dk, \qquad R = \Biggl(\frac{3M}{4 \pi \bar\rho_{m}}\Biggr)^{\frac{1}{3}}. 
\end{equation}
The smoothing function $W$ is the Fourier transform of a top-hat filter (the same functional form of \eqref{tophat_filter}).
The function $f(\sigma)$ is generally calibrated from N-body simulation.
In this work we use a Tinker mass function~\cite{Tinker:2008}.
Within this framework, the shot-noise and the HI bias can be written as \cite{Bull2015, Castorina:2016}
\begin{align}
N_\text{HI}(z) &= \Biggl(\frac{1}{\bar\rho_\text{HI}(z)}\Biggr)^2 \int^{\infty}_0 n(M,z) M^2_\text{HI}(z) dM, \label{NHI}\\
b_\text{HI}(z) &=  \frac{1}{\bar\rho_\text{HI}(z)} \int^{\infty}_0 n(M,z) b(M,z) M_\text{HI}(z) dM, \label{bHI}
\end{align}
where $b(M,z)$ is the halo bias, calibrated on the N-body simulation from \cite{Tinker:2010}.

We model the average HI mass within an halo of mass $M$ as redshift independent \cite{Castorina:2016}
\begin{equation}
M_\text{HI}(M, z) = \text{C} \  (1-Y_\text{p}) \frac{\Omega_\text{b}}{\Omega_\text{m}} \exp\Bigl[{-(M_\text{min}/M)} \Bigr] M^{\alpha},
\label{model}
\end{equation}
where $Y_\text{p}=0.24$ is the Helium fraction (note that $ (1-Y_\text{p}) \frac{\Omega_\text{b}}{\Omega_\text{m}}$ is the HI mass fraction), $M_\text{min}$, $\alpha$ and C are the free parameters of the model. 
Eq.~\eqref{model} entails that the mass of cosmic hydrogen within a halo scales as a power law of the total virial mass of the halo, at large halo masses. The efficiency of this scaling is regulated by the exponent $\alpha$ (larger values of $\alpha$ correspond to larger amount of 
hydrogen within a halo of fixed mass).
At low halo masses, we expect instead an exponential suppression, due to different physical processes, such as photoionization from
the UV background or galactic winds. The parameter $M_\text{min}$ regulates the range of halo masses for which this suppression
is effective (larger values of $M_\text{min}$ imply a larger range of masses for which the suppression is relevant).
The parameter ${\rm C}$ is an overall normalization constant, that needs to be fixed by matching the theoretical abundance of neutral hydrogen,
predicted by \eqref{OmHI}, with what is measured by HI galaxy survey at a given redshift.
As we can notice from Eqs.~\eqref{NHI} and~\eqref{bHI}, both the shot-noise and the HI bias do not depend on this normalization factor, therefore we can fix $\text{C} = 1$ without loss of generality.

\begin{figure}
\centering
    \begin{subfigure}[b]{0.485\textwidth}
        \includegraphics[width=\textwidth]{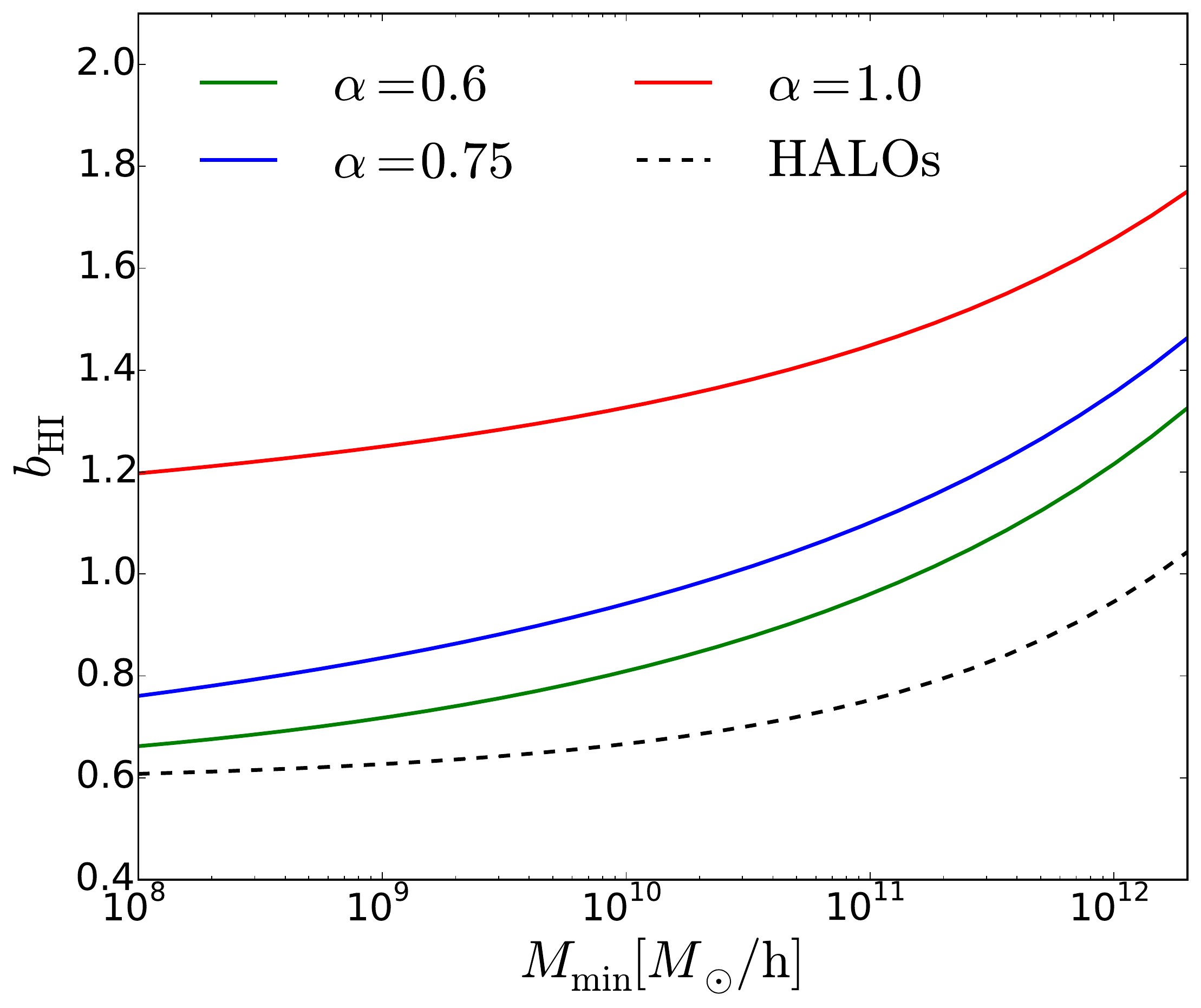}
        \caption{HI bias}
        \label{fig:bias}
    \end{subfigure}
    \begin{subfigure}[b]{0.5\textwidth}
        \includegraphics[width=\textwidth]{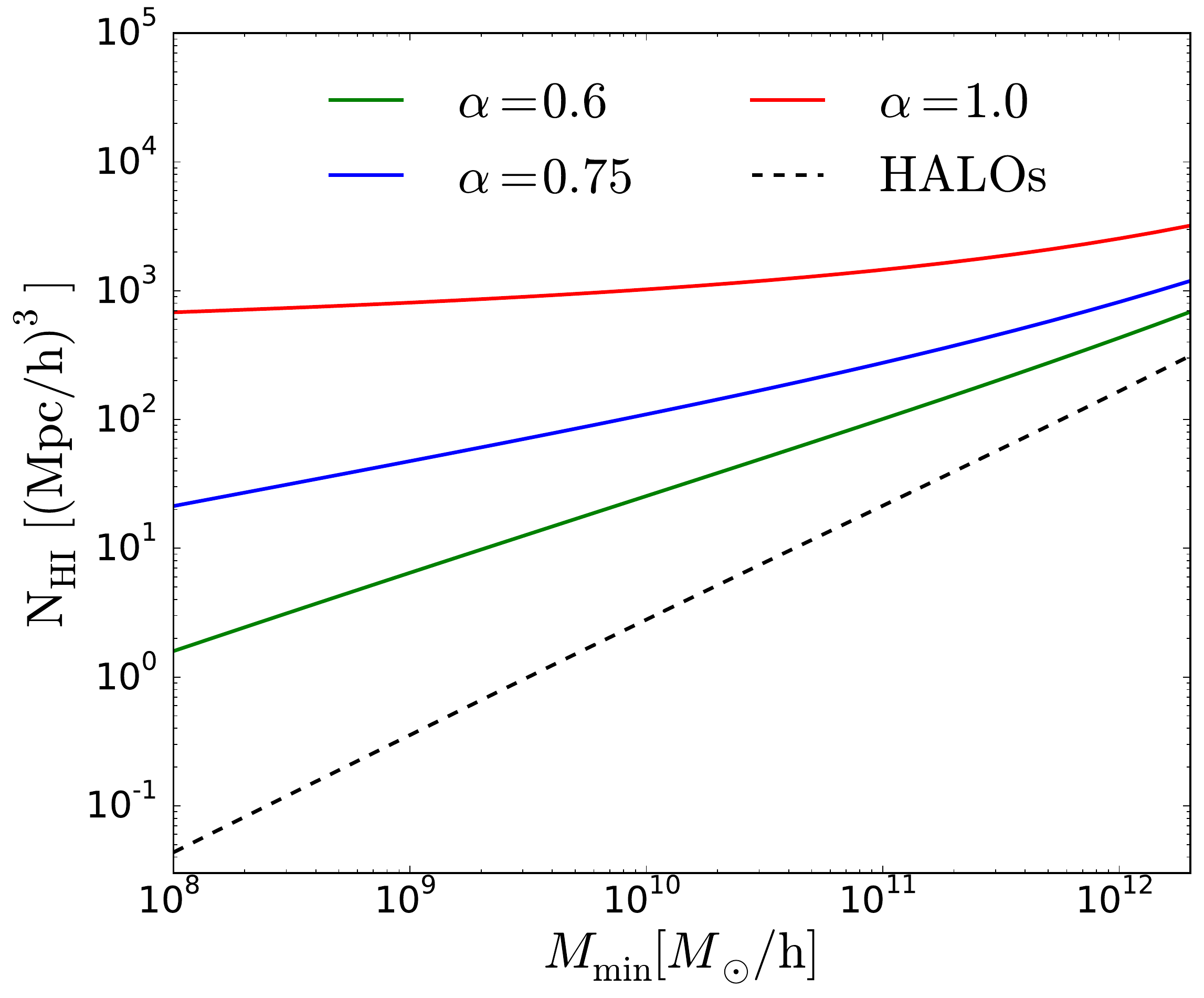}
        \caption{HI shot-noise}
        \label{fig:noise}
    \end{subfigure}
    \caption{HI bias (left panel) and HI shot-noise (right panel) as a function of the cutoff parameter $M_\text{min}$, at redshift
$z = 0.15$.
Different colors denote different values of the exponent $\alpha$ in \eqref{model}. The black dashed lines represent the 
halo bias and the shot-noise of the halo population.}
    \label{fig:HI}
\end{figure}

\begin{figure}
\centering
    \begin{subfigure}[b]{0.495\textwidth}
        \includegraphics[width=\textwidth]{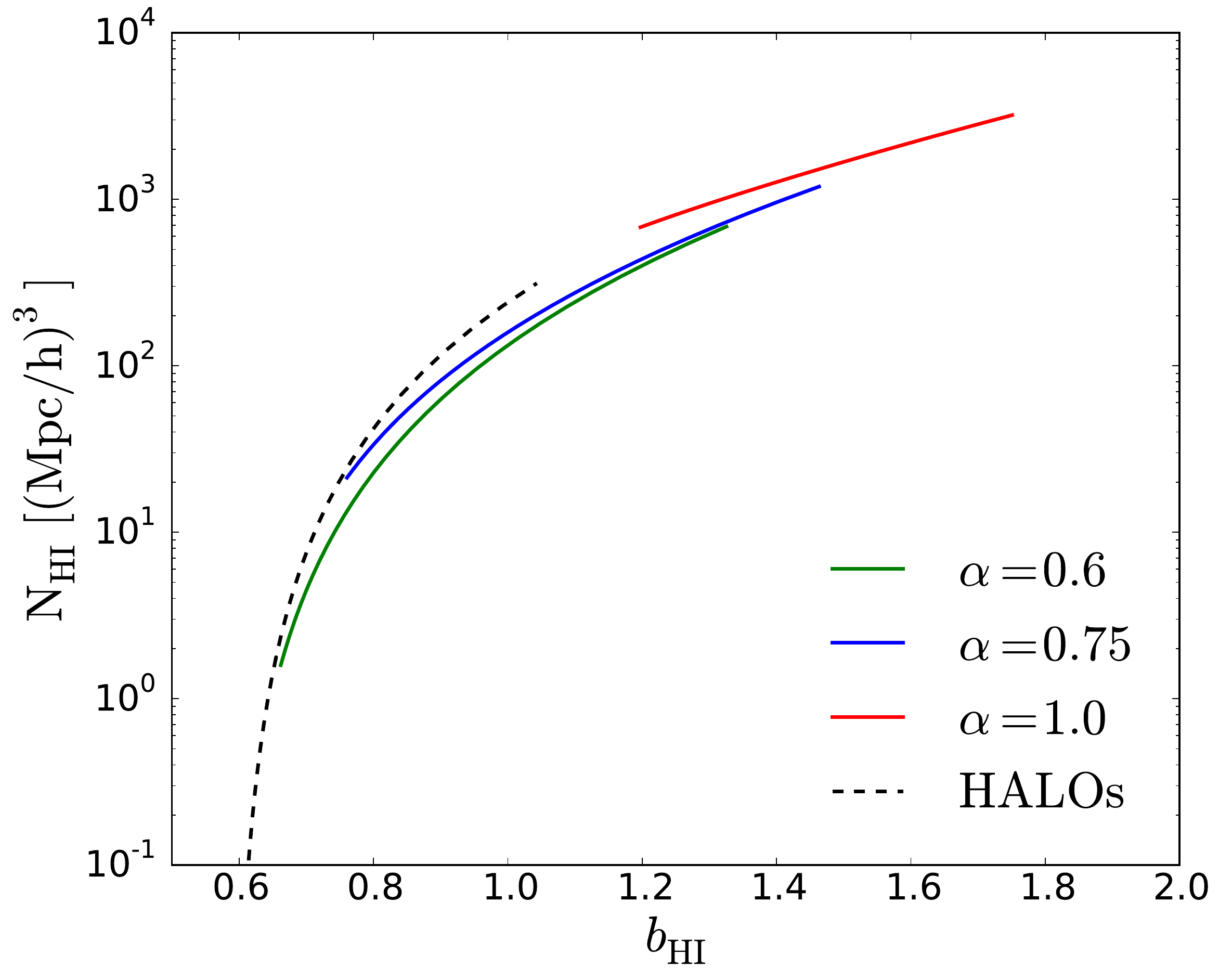}
        \caption{Shot noise VS bias - HI}
        \label{SN_vs_bias}
    \end{subfigure}
    \begin{subfigure}[b]{0.495\textwidth}
        \includegraphics[width=\textwidth]{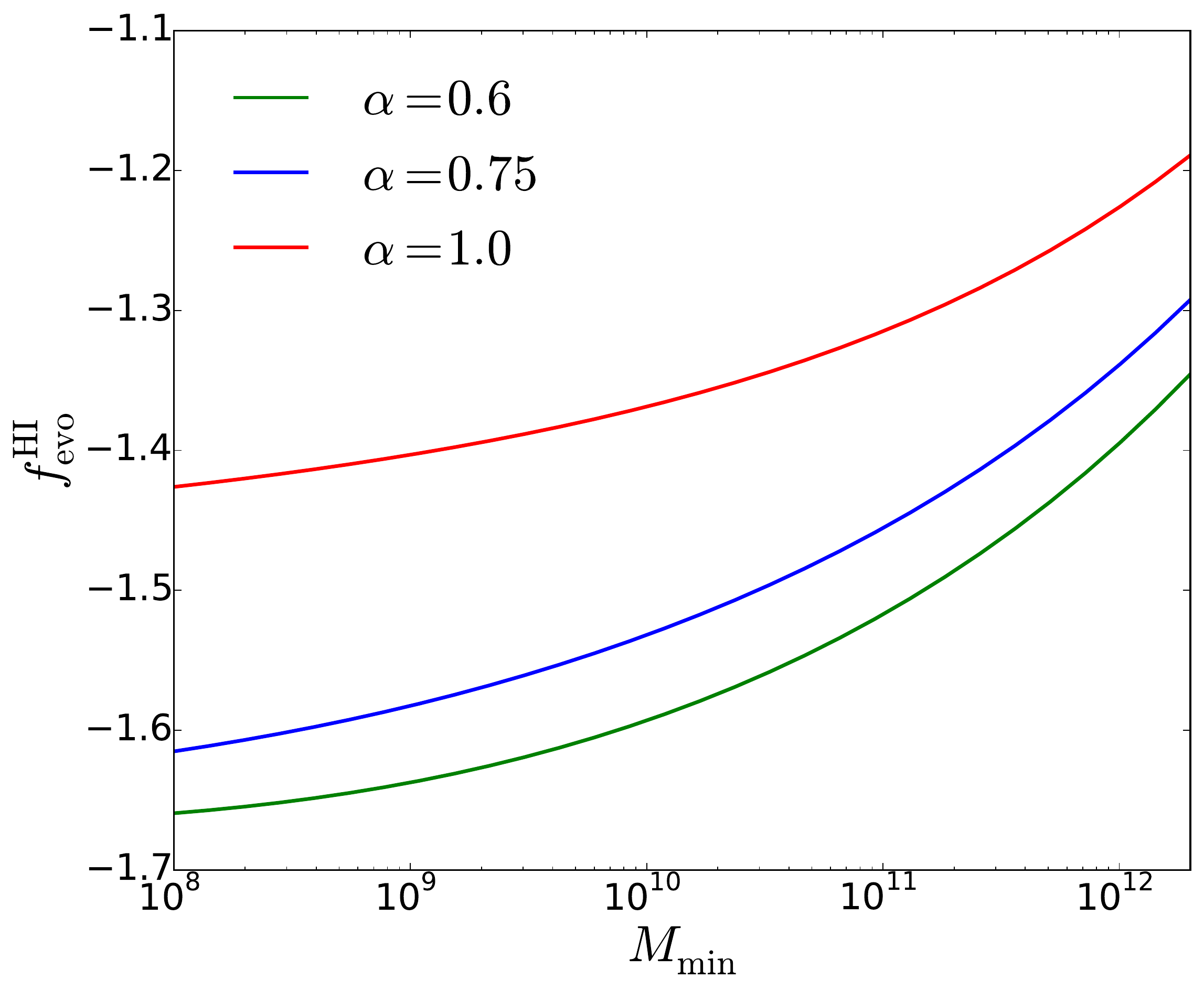}
        \caption{Evolution bias - HI}
        \label{fig:fevo}
    \end{subfigure}
    \caption{Left panel: Shot-noise of the HI as a function of the HI bias. 
    Right panel: Evolution bias as a function of the cutoff parameter $M_\text{min}$. 
     The notation is the same as figure \ref{fig:HI}.}
\end{figure}

In figure~\ref{fig:HI} we show the HI bias and HI shot-noise, respectively, as a function
of the cutoff parameter $M_\text{min}$, for different values of $\alpha$.
We can see that if the neutral hydrogen is concentrated in halos with large mass (this corresponds to larger values for $M_\text{min}$ and $\alpha$), both the shot-noise and the HI bias have larger values with respect to the case in which we find considerable amount of 
neutral hydrogen also in small halos.
The halo bias (in black dashed lines) is computed from Eq.~(15) in \cite{Marulli:2011}
\begin{equation}
b_\text{halos}(z)  = \frac{\int_0^{\infty} n(M, z) b(M, z) dM}{\int_0^{\infty} n(M,z) dM},
\end{equation}
while the shot-noise for the halo population is computed as
\begin{equation}
N_\text{halos} = \frac{1}{\int_0^{\infty} n(M, z) dM}.
\end{equation}
From figure~\ref{fig:HI} we can find a relation between the bias of HI, $b_{\rm HI}$, and the shot-noise, $N_{\rm HI}$.
In figure \eqref{SN_vs_bias} we show how the HI bias and the HI shot-noise are related to each others, for different values of
the parameter $\alpha$, by considering $M_{\rm min}$ as the parametrisation of the different curves.
As expected, in all models higher biased HI models correspond to higher shot-noise.
Furthermore, models with an higher efficiency in accreting neutral hydrogen within halos (larger values of $\alpha$) correspond 
to higher values of bias and shot-noise.
According to hydrodynamics simulations~\cite{Paco:2015a, Paco:2015b} and semi-analytic models~\cite{Kim:2017},
values of $\alpha < 0.9$ are more realistic.

In figure \ref{fig:fevo} we show the dependence of the evolution bias, computed from \eqref{fevo_HI}, for different models.
The evolution bias for the HI does not vary significantly within the parameter space, because it is described by the redshift evolution of the halo mass function. Its negative value indicates that the density of HI drops at low redshift.

We note that the model outlined in this section is more focussed on the high redshift regime and needs to be better tested
and improved in the low redshift Universe in order to capture the complex astrophysical effects \cite{evoli11,mancuso17} that
could impact on the HI distribution inside galaxies. 

\subsection{Galaxy model}
\label{Sec:gal}
As discussed in the previous section for neutral hydrogen, we can use the halo model to relate the bias parameters and the shot-noise of galaxy distributions.
In this framework, we assume that all the galaxies are found within dark matter halos. The relevant quantity
we need to model is their comoving number density, which can be computed from the halo mass function as
\begin{equation}
\label{ng}
n_\text{g}(z) = \int^{\infty}_{0} n(M, z) N_\text{av}(M,z) dM,
\end{equation}
where $N_\text{av}$ is the average number of galaxies for an halo of mass $M$.
The galaxy bias can be modeled as 
\begin{equation}
b_\text{g} (z) =  \frac{1}{n_\text{g}(z)} \int^{\infty}_0 n(M,z) b(M,z) N_\text{av}(M,z) dM, \label{b_gal}
\end{equation}
while the shot-noise is simply given by the inverse of the number density, i.e.
\begin{equation}
N_\text{gal} \equiv \frac{1}{n_\text{g}} = \dfrac{1}{\int^{\infty}_0 n(M, z)  N_\text{av}(M,z) \,dM}. \label{noise_gal}
\end{equation}
We model the average number of galaxies within a halo of mass $M$ as
\begin{equation}
\label{Nav}
 N_\text{av}(M) =
\left \{
  \begin{tabular}{lc}
$0$ \quad &if $M \le M^*_\text{min}$ \\
 & \\
$A  \Bigl(\frac{M}{M^*_\text{min}}\Bigr)^{\alpha_\text{gal}}$  \quad &if $M > M^*_\text{min}$
  \end{tabular}
\right .
\end{equation}
This model is similar to the model employed in \cite{cooray:2002} to model red and blue galaxies, which provides a good fit to 
the number of subhalos expected from numerical simulations \cite{White:2000}.
The parameter $M^*_\text{min}$ represents a threshold mass, below which a halo cannot host a galaxy, because their potential wells are shallower
with respect to more massive halos. Therefore, some physical processes such as supernova feedback, can be efficient enough
to expel a huge percent of baryons from the halo and therefore suppress the star formation within the halo itself.
For halos with masses larger than this threshold, we assume that the average number of galaxies per halo increases with the halo mass, following a power law $N_\text{av}(M) \propto M^{\alpha_\text{gal}}$, so that $\alpha_\text{gal}$ represents its slope.
The parameter $A$ represents a normalization constant. 
From Eq.~\eqref{b_gal} we see that, as we found for the HI, the galaxy bias is not affected by the value of the normalization, 
while the shot-noise, Eq.~\eqref{noise_gal}, strongly depends on $A$.
In the literature more flexible and physically motivated models have been proposed (see e.g. \cite{Zheng:2004, Yang:2011wa}). Nevertheless, the simple model described above captures the features that we need for the purpose of our analysis with the minimal number of parameters.

\begin{figure}
\centering
    \begin{subfigure}[b]{0.485\textwidth}
        \includegraphics[width=\textwidth]{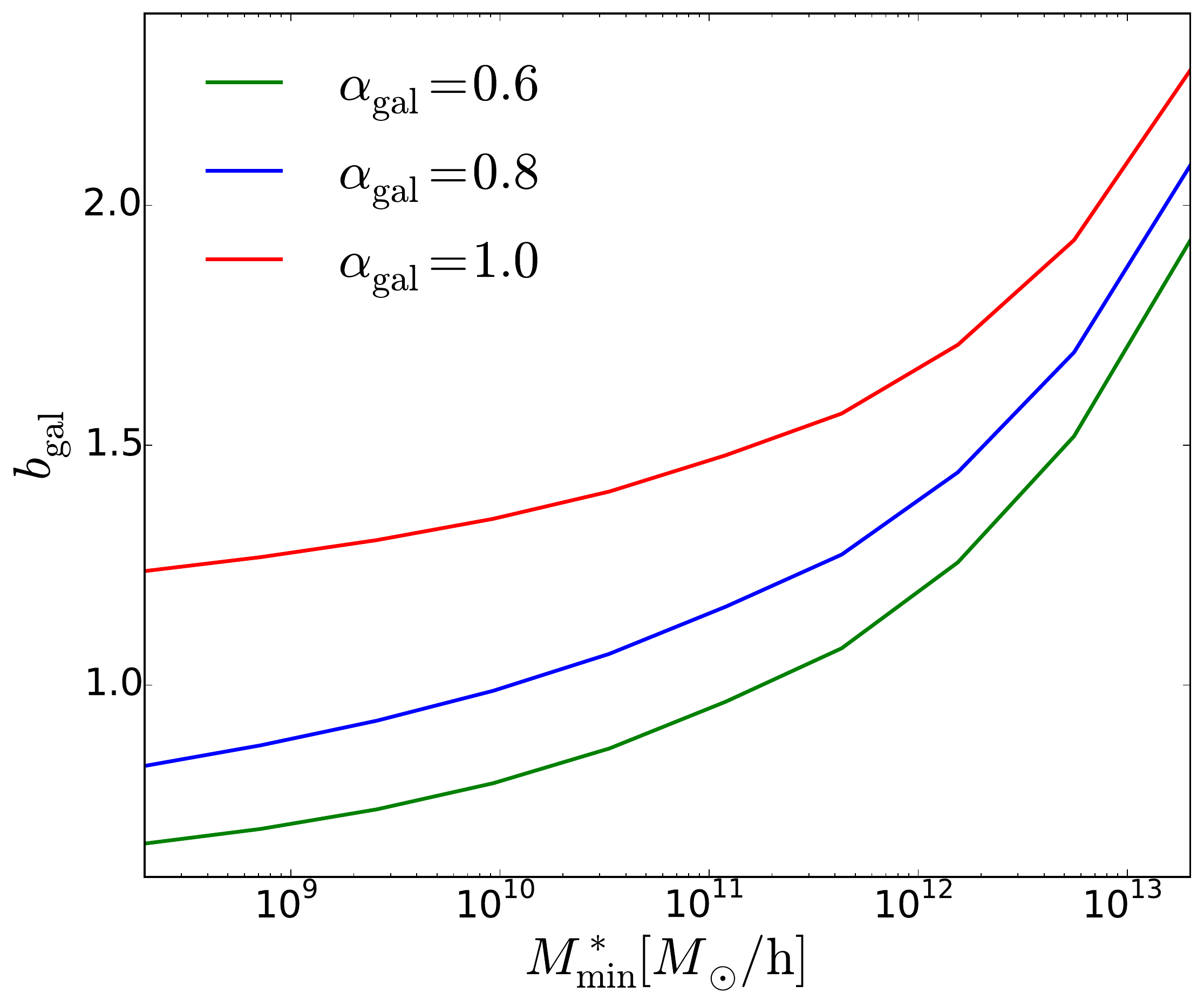}
        \caption{Galaxy bias}
        \label{fig:bias_gal}
    \end{subfigure}
    \begin{subfigure}[b]{0.5\textwidth}
        \includegraphics[width=\textwidth]{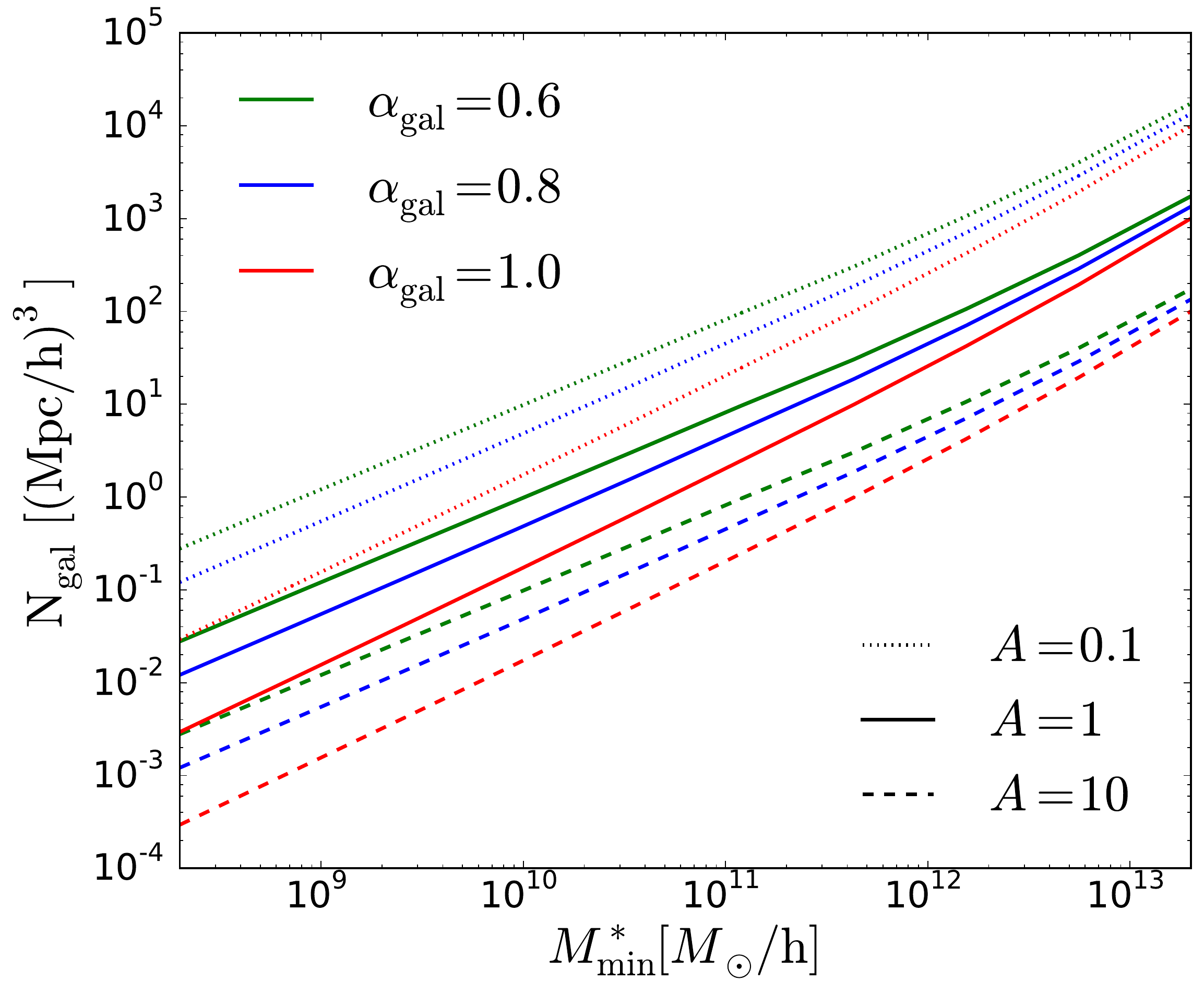}
        \caption{Galaxy shot-noise}
        \label{fig:noise_gal}
    \end{subfigure}
    \caption{Galaxy bias (left panel) and galaxy shot-noise (right panel) as a function of the cutoff parameter $M^*_\text{min}$, at redshift
$z = 0.15$.
Different colors denote different values of the exponent $\alpha_\text{gal}$ in \eqref{model}. 
In the right panel, different line-styles denote different values of the normalization constant $A$.
}
    \label{fig:gal}
\end{figure}

In figure \eqref{fig:gal} we show the galaxy bias (left panel) and galaxy shot-noise (right panel) as a function of the
threshold mass $M^*_\text{min}$.  
The bias increases exponentially with the threshold mass, while the shot-noise increase as a power law whose slope depends 
the slope of the model \eqref{Nav}.

\begin{figure}[t]
\begin{center}
\includegraphics[width=0.8\textwidth]{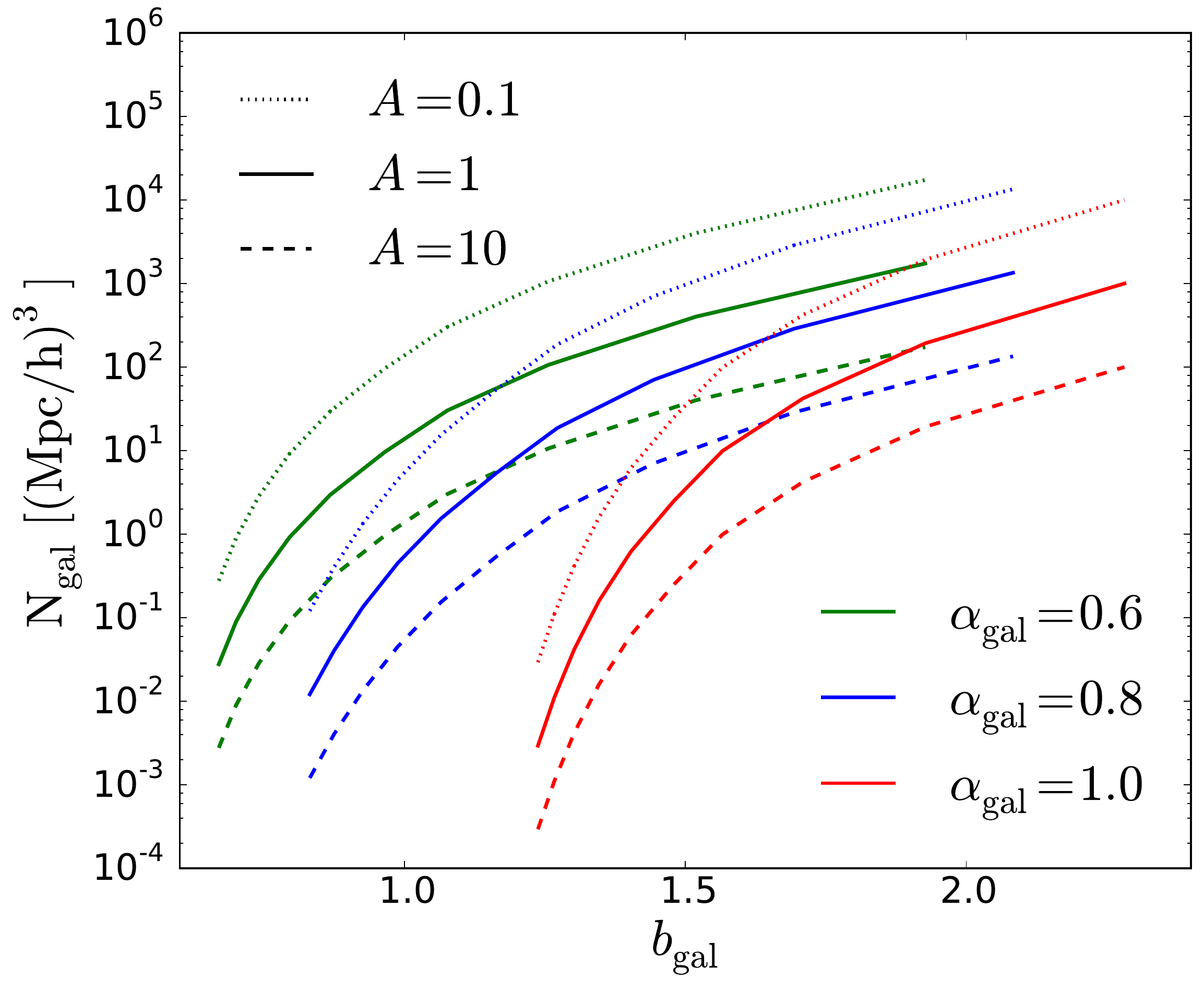}
\caption{Shot-noise of the galaxies as a function of the galaxy bias, for different values of the slope $\alpha_\text{gal}$
and different values of the normalization constant $A$.
}
\label{noise_vs_bias_gal}
\end{center}
\end{figure}

In figure \eqref{noise_vs_bias_gal} we summarize the information of figure  \eqref{fig:gal} by displaying the
relation between galaxy bias and shot-noise, for different models.
Similarly to HI, highly biased galaxy populations (which are more massive) show a larger shot-noise,
compared to the one with lower bias.

\section{Signal-to-noise analysis}
\label{Sec:SN}
In this section we will present a signal-to-noise analysis for the dipole of the HI-galaxies cross-correlation.
The signal-to-noise for the dipole is defined as \cite{Hall:2016}
\begin{equation}
\left(\frac{S}{N}\right)^2 = \sum^{d_\text{max}-L_\text{p}/2}_{d_1, d_2 = d_\text{min}+L_\text{p}/2} \xi_1(d_1)  \mbox{COV}^{-1}(d_1, d_2)  \xi_1(d_2)\, . \label{SN}
\end{equation}
The minimum distance $d_\text{min}$ is chosen to be the non-linear scale. 
We set $d_\text{min} = 30\, \text{Mpc/}h$, which
correspond to $k_\text{NL} \approx 0.2 \,h/\text{Mpc}$. We also study the dependence of the signal-to-noise ratio on both $d_{\rm min}$ and $d_{\rm max}$. 
\begin{figure}
\centering
    \begin{subfigure}[b]{0.49\textwidth}
        \includegraphics[width=\textwidth]{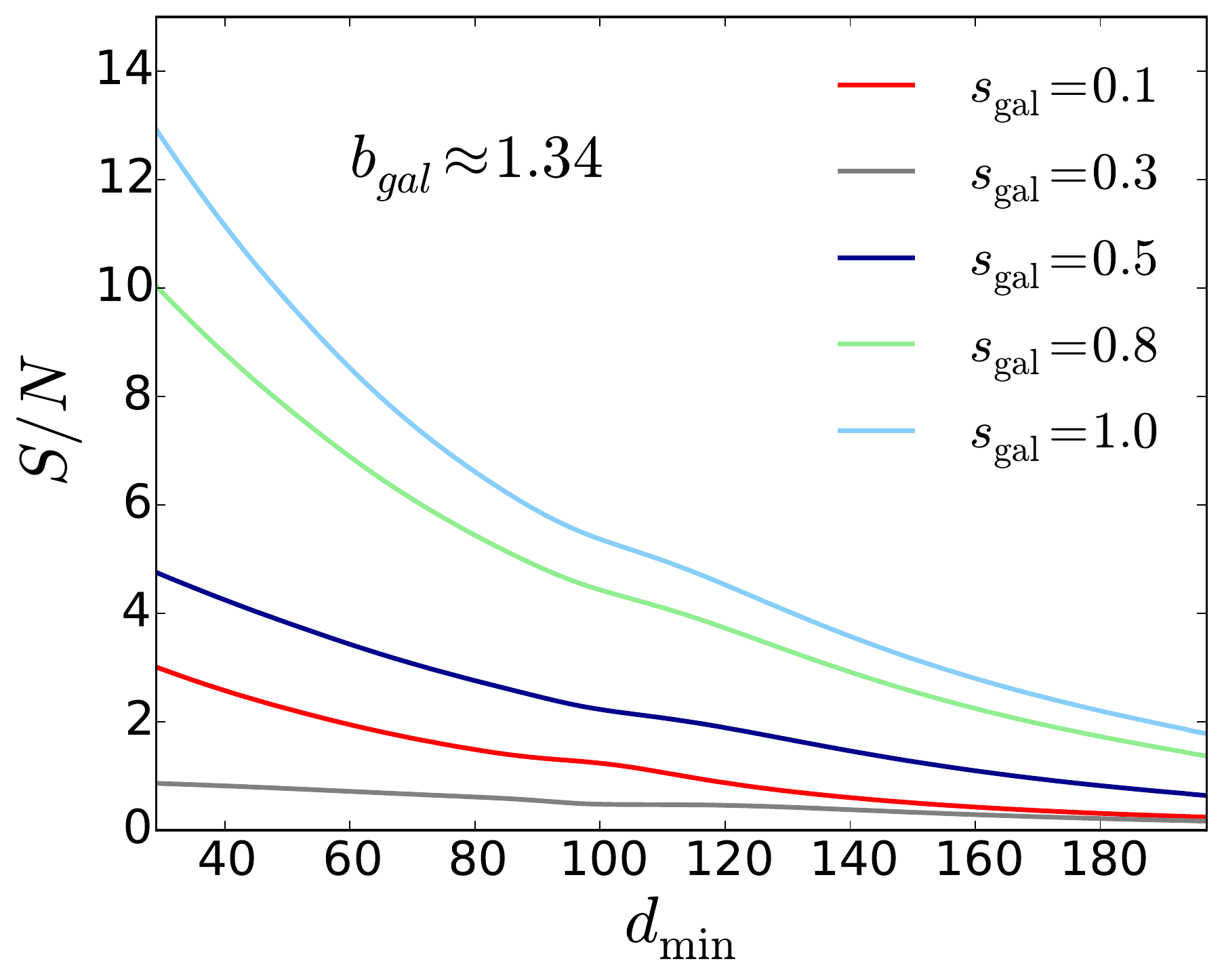}
    \end{subfigure}
    \begin{subfigure}[b]{0.49\textwidth}
        \includegraphics[width=\textwidth]{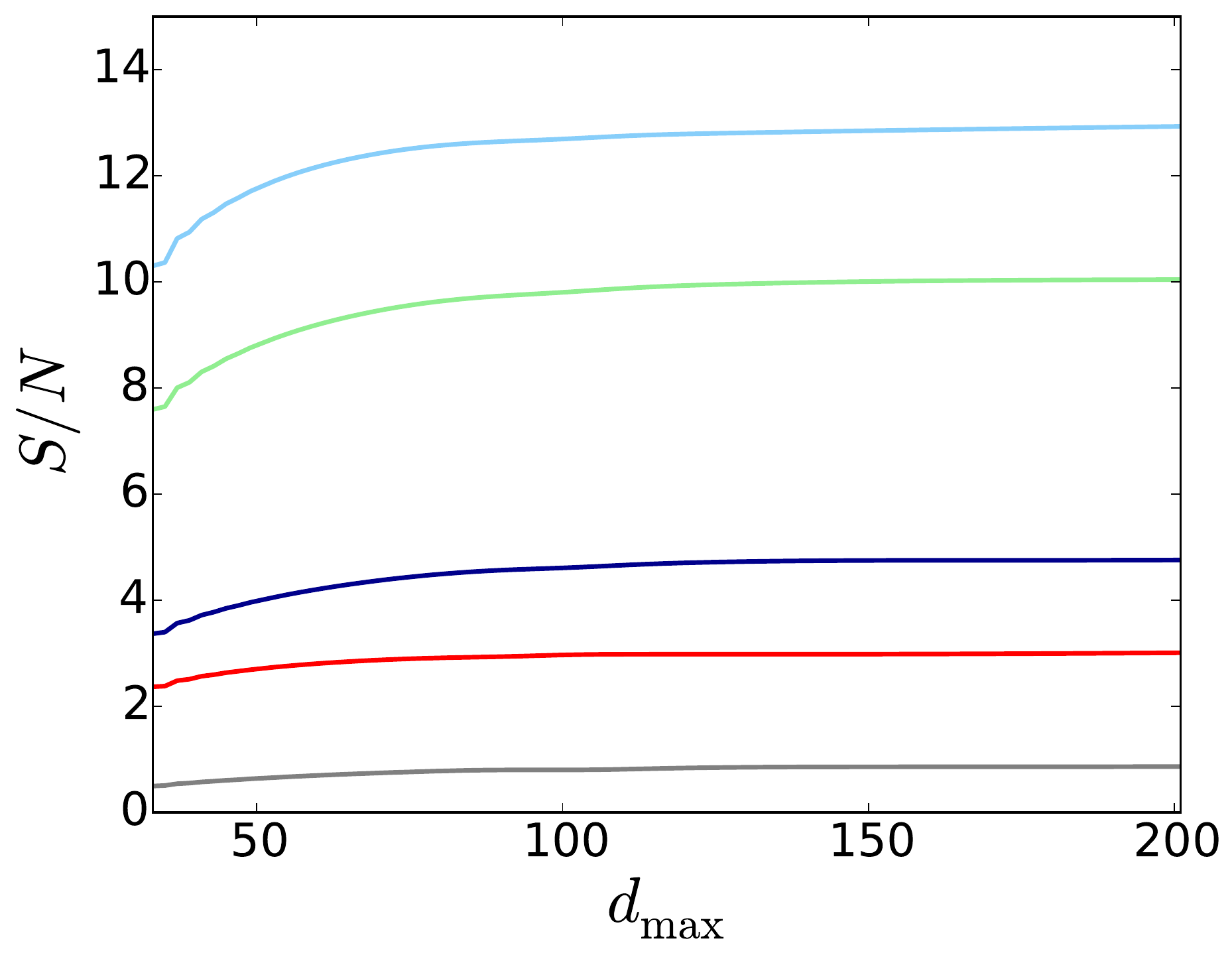}
    \end{subfigure}
    \caption{Signal-to-noise as a function of $d_\text{min}$ and $d_\text{max}$.
    The HI model is the model 1 in Table \ref{table_HImod}. 
    The galaxy bias is fixed to be $b_\text{gal} \approx 1.34$ here, and the interferometer noise is set to be $P_\text{N} = 100\, (\text{Mpc}/{h})^3$.
    Different colors denotes different values for the magnification bias.
}
    \label{fig:dmin_dmax}
\end{figure}
The maximum distance is set to be the distance at which the dipole estimator, computed to lowest order in the $d/r$ expansion from \eqref{wa_corr}, 
coincides with the full-sky quantity up to $3\%$. 
Since we are interested in relatively low redshift measurements, we provide a comparison between the wide-angle correction to linear order in $d/r$ and the exact full-sky contributions.
The details of this comparison are shown in the Appendix \ref{Ap_A}.  
Our results show that for $d \le 200 \, \mathrm{Mpc}/h$ the discrepancy between the two quantities is
smaller than the $3\%$ threshold, therefore we set $d_\text{max} = 200 \, \mathrm{Mpc}/{h}$. 
Although more numerically expensive, it is possible to consider the full-sky expression if further accuracy is required. Nevertheless, as shown in figure \ref{fig:dmin_dmax}, the main information is encoded on scales where the expansion to linear order in $d/r$ is accurate enough.

The cross-correlation volume $V$ is computed as
\begin{equation}
V = \frac{4}{3}\pi  \Bigl(r^3(z_\text{max}) - r^3(z_\text{min})\Bigr) f_\text{sky},
\end{equation}
where $r(z)$ is the comoving distance at redshift $z$, $z_\text{max}$ and $z_\text{min}$ are the maximum and minimum observed redshift,
respectively, and $f_\text{sky}$ is the fraction of the sky available for the observation.
We consider a redshift range between $z_\text{min} = 0.05$ and $z_\text{max} = 0.25$ and a fraction of sky
$f_\text{sky} = 0.2$, which is the sky coverage of a survey similar to the Baryon Oscillation Spectroscopic Survey (BOSS) \cite{Alam:2016hwk}. 
We assumed the interferometer employed for observing the neutral hydrogen in intensity mapping can resolve pixels of size
$L_\text{p} = 2 \,\text{Mpc}/{h}$.

We will consider two models for the neutral hydrogen, based on the formalism presented in section \ref{HI_sec}.
The first model (conservative model) is similar to the one employed in \cite{Paco:2016}, with bias close to unity at low redshift
and relatively high shot-noise.
The second model (optimistic model) is similar to the one adopted for the forecast in \cite{Bull2015}, with smaller
bias and lower shot-noise.
The specific values of the parameters of the two models are highlighted in Table \ref{table_HImod}.

Concerning the galaxy model, the parameter A is a normalization constant and does not qualitatively affect the relations between
shot-noise and bias. Therefore, in the rest of the paper we will fix its value to be $A = 1$, which correspond to the minimum value
of the average number of galaxies per halos whose masses are above the threshold value  $M^*_\text{min}$.
Furthermore, we fix the value of $\alpha_\text{gal} = 1$, which corresponds to the most biased galaxies. 
\begin{table}[tbp]
\centering
\begin{tabular}{|c|c|c|}
\hline
\multicolumn{1}{|c|}{}                    & 
 \multicolumn{1}{c|}{MODEL 1}&
  \multicolumn{1}{c|}{MODEL 2} \\
\multicolumn{1}{|c|}{}                    & 
 \multicolumn{1}{c|}{(CONSERVATIVE)}&
  \multicolumn{1}{c|}{(OPTIMISTIC)} \\
\hline
$\alpha$                        &    0.75                                                                      &  0.6                                                    \\
$M_\text{min}$              &    $\approx 1.7 \times 10^{10} M_{\odot}/\text{h}$   &   $10^8 M_{\odot}/\text{h}$                \\
$b_\text{HI}$                 &     $\approx 0.99$                                                    &    $\approx 0.67$                                  \\
$N_\text{HI}$                &    $\approx 143 \text{Mpc}^3/\text{h}^3$                  &    $\approx 2 \text{Mpc}^3/\text{h}^3$   \\
$f^\text{HI}_\text{evo}$ &    $\approx - 4.33$                                                   &   $\approx - 4.40$                                   \\
\hline
\end{tabular}
\caption{
Values of the parameters for the two HI models we consider for our analysis. 
Model 1 has larger bias and larger shot-noise with respect to Model 2.
}
\label{table_HImod}
\end{table}

\begin{figure}[t]
\centering
    \begin{subfigure}[b]{0.495\textwidth}
        \includegraphics[width=\textwidth]{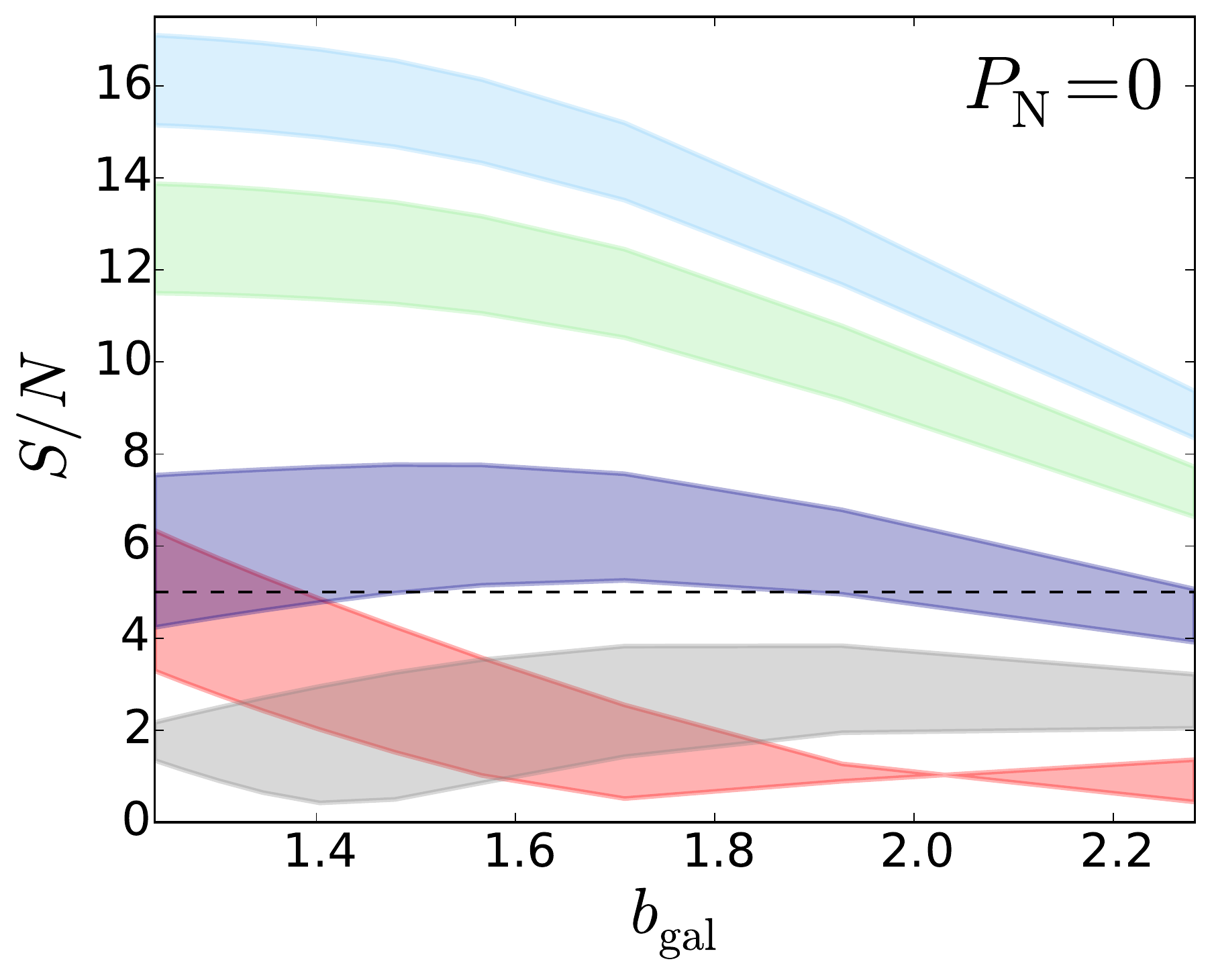}
        \caption{MODEL 1 (CONSERVATIVE)}
        \label{fig:mod1}
    \end{subfigure}
    \begin{subfigure}[b]{0.495\textwidth}
        \includegraphics[width=\textwidth]{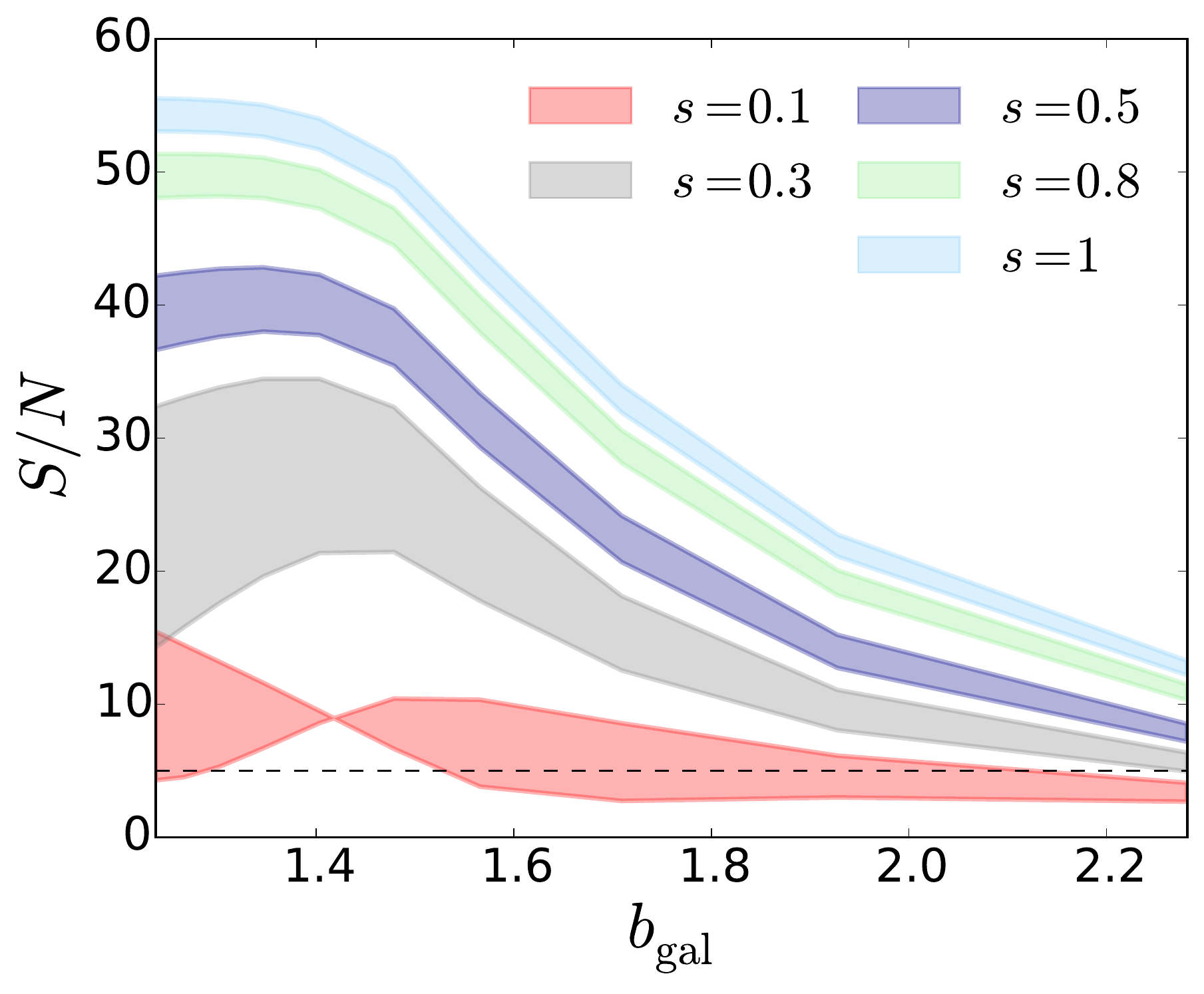}
        \caption{MODEL 2 (OPTIMISTIC)}
        \label{fig:mod2}
    \end{subfigure}
    \caption{Signal-to-noise for the cross-correlation dipole as a function of the galaxy bias.
The slope of the model for the galaxy population is fixed at the value $\alpha_\text{gal} = 1$.
Different colors denote different values of the magnification bias. The corresponding shaded region represents the signal-to-noise for $f^{gal}_\text{evo} \in [-2, 2]$. The interferometer noise is neglected here. 
}
    \label{fig:SN_bias_alp1}
\end{figure}

\begin{figure}[t]
\centering
    \begin{subfigure}[b]{0.495\textwidth}
        \includegraphics[width=\textwidth]{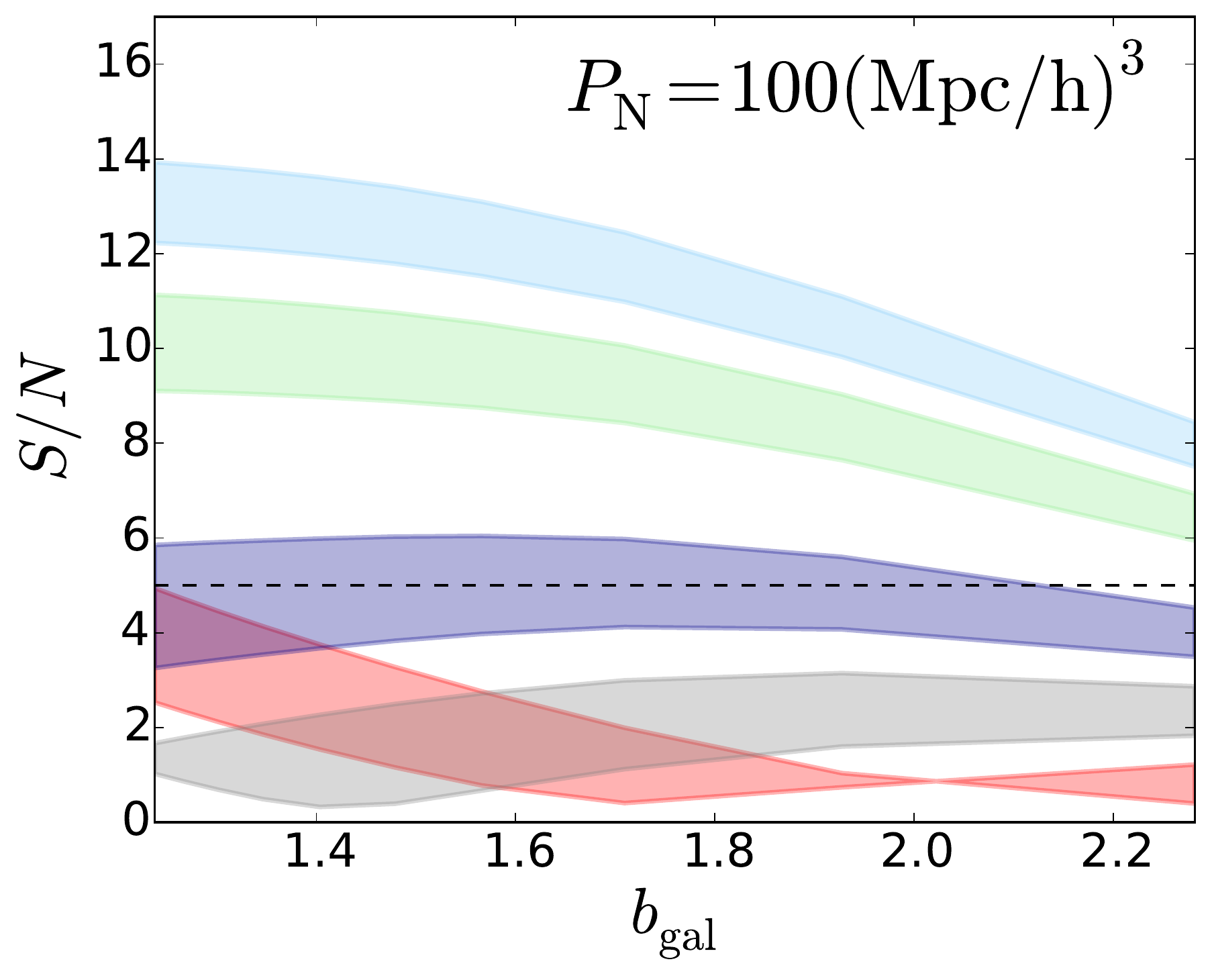}
        \caption{MODEL 1 (CONSERVATIVE)}
        \label{fig:mod1_noise}
    \end{subfigure}
    \begin{subfigure}[b]{0.495\textwidth}
        \includegraphics[width=\textwidth]{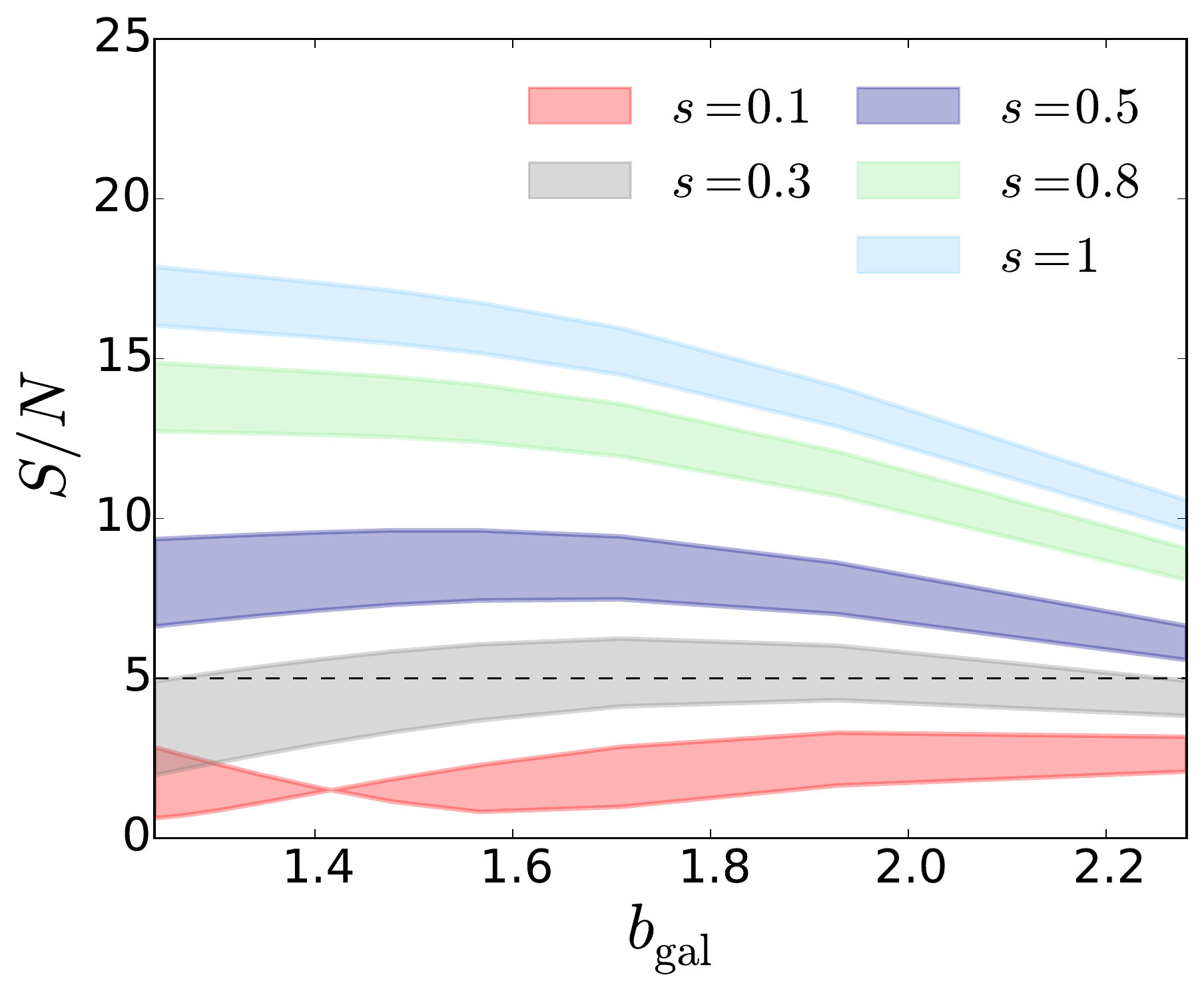}
        \caption{MODEL 2 (OPTIMISTIC)}
        \label{fig:mod2_noise}
    \end{subfigure}
    \caption{Signal-to-noise for the cross-correlation dipole as a function of the galaxy bias, with an interferometer noise
    $P_\text{N} = 100 \, (\text{Mpc}/\text{h})^3$. The same conventions of figure \ref{fig:SN_bias_alp1} are adopted.
    The black dashed line highlight the $S/N = 5$ threshold. 
}
    \label{fig:SN_bias_alp1_noise}
\end{figure}

Figure \ref{fig:SN_bias_alp1} shows the dipole signal-to-noise at $z = 0.15$, as a function of the galaxy bias, for the two
models described above. 
The signal-to-noise is computed for different values of the magnification
bias, which is treated here as a free parameter. 
We find that the signal-to-noise is generally optimized for the largest value of the magnification bias (which in this case is $s_\text{gal} = 1$). Let us remark that the signal-to-noise ratio does not grow monotonically with the magnification bias, but it has a minimum between the lines $s=0.1$ and $s=0.3$ of Fig.~\eqref{fig:SN_bias_alp1}. Indeed as shown in Fig.~\eqref{Fig_WA}, the largest pre-factor of the dipole in flat-sky is $(2-5 s)/( r\HH)$.
Then, considering wide-angle effects and the impact of the other bias factors (in particular the large negative value of the evolution bias of neutral hydrogen), the dipole vanishes for lower values of magnification bias. 

In the optimistic model for the HI (figure \ref{fig:mod2}), the signal-to-noise is $> 5$ for all the values of the magnification bias. 
We see that it increases with the bias up to $b_\text{gal} \approx 1.4$, then it declines for more biased galaxies. 
This is due to the fact that the dipole signal increases for larger difference between the biases of the two tracers, but highly
biased tracers are more massive and we observe fewer of them. Therefore for the most biased galaxies the growth of the shot-noise dominates
over growth of the signal and the $S/N$ results to be suppressed.

Similar comments are valid for the more conservative HI model. In this case we need $s_\text{gal} \ge 0.5$ in order to have a 
$S/N > 5$.
The maximum value of the signal-to-noise in the two models ranges between $(S/N)_\text{max}\approx 17$, for the more conservative model, to $(S/N)_\text{max}\approx 53$ in the optimistic model.

In figure \ref{fig:SN_bias_alp1_noise} we show how the interferometer noise affects the signal-to-noise analysis for the two models.
We assume the fiducial value $P_\text{N} = 100 \, (\text{Mpc}/{h})^3$, which corresponds to the noise of a CHIME-like survey 
\cite{CHIME, Bandura:2014gwa} observing at the mean redshift considered in our work, computed from Refs.~\cite{Hall:2016, Bull2015}.
For both models the signal-to-noise is suppressed and the effect results to be more prominent for the optimistic model.
We see that the interferometer noise suppresses the maximum value of the signal-to-noise ratio roughly from $17$ to $ 14$ for the conservative model, while the maximum $S/N$ decreases from $ 53$ to $ 18$ in the optimistic model.
Indeed, in the optimistic model the HI shot-noise results to be much smaller than the reference interferometer noise, while in the conservative model
the two quantities are comparable. 
Interestingly, we observe that, even when the interferometer noise is included in the analysis, 
the signal-to-noise results to be $> 5$ for  $s_\text{gal} \ge 0.5$ in both models we considered for the HI. 
 
\section{HOD approach for modelling Luminosity-threshold galaxy catalogues}
\label{Sec:HOD}
The galaxy model described in section \ref{Sec:gal}, and applied in the signal-to-noise analysis in section \ref{Sec:SN},
is intuitive and easy to implement. 
Nevertheless, its limitations are many. In fact, galaxy surveys can observe sources with luminosity
larger that a threshold value. Within the framework described in \ref{Sec:gal} , it is not possible to model the luminosity function of 
a sample of galaxies, and therefore both magnification and evolution bias were treated as free parameters.
In this section we will partially trade the generality of this framework with a more realistic model for a galaxy catalogue.

\begin{figure}[t]
\centering
    \begin{subfigure}[b]{0.495\textwidth}
        \includegraphics[width=1.0\textwidth]{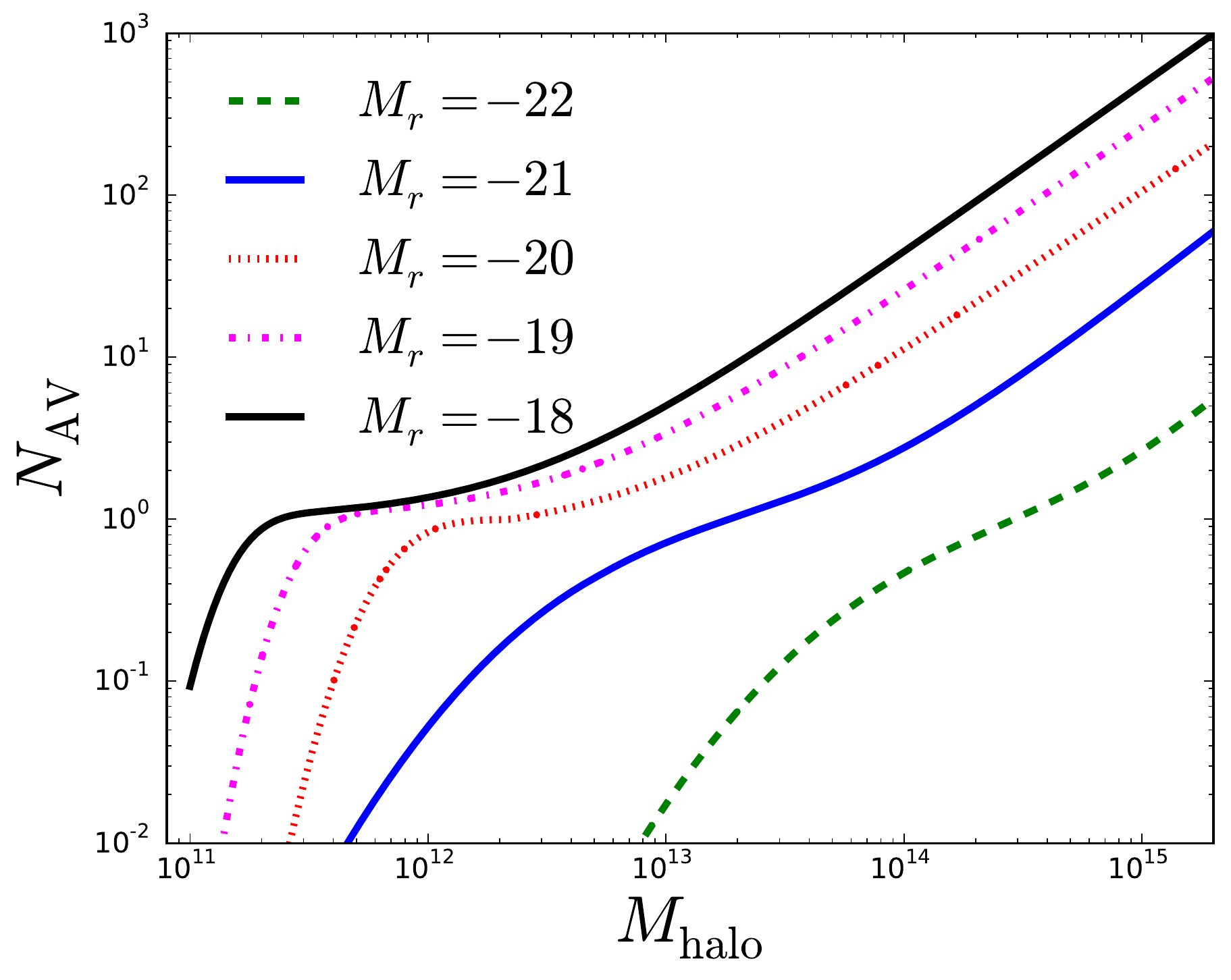}
        \label{HOD_Ngal}
    \end{subfigure}
    \begin{subfigure}[b]{0.495\textwidth}
        \includegraphics[width=1.06\textwidth]{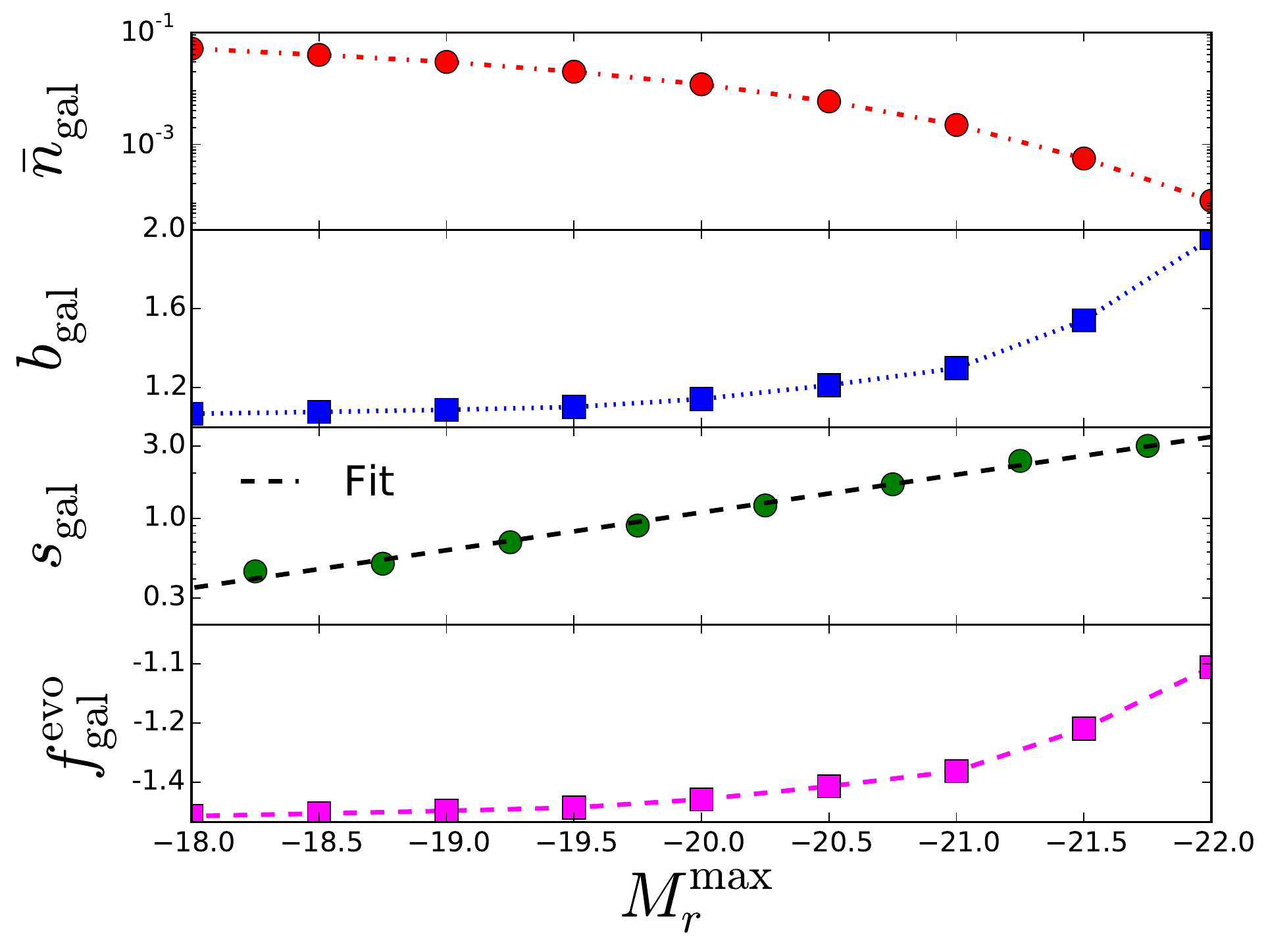}
        \label{HOD_models}
    \end{subfigure}
    \caption{Left panel: Average number of galaxies within a halo as a function of the halo mass. 
Different lines denote galaxies samples with different values of the maximum absolute magnitude $M_r$.
Right panel: Comoving number density (red), clustering bias (blue), magnification bias (green) and evolution bias (magenta) 
as a function of the maximum absolute magnitude $M_r$ of the selected galaxy sample. 
}
    \label{fig:HOD}
\end{figure}

In order to model a luminosity-threshold galaxy catalogue, we will assume the model based on the Halo Occupation
Distribution (HOD) described in \cite{Zheng:2007, Zehavi:2011}. 
The average number of galaxies, within a halo of mass $M$ and with apparent magnitude below a certain threshold value $m^*$ is modelled as a sum of two contributions: the contributions from the central galaxies and the one from
the satellite galaxies \cite{Kravtsov:2003, Zheng:2004}
\begin{equation}
N_\text{AV}(< m^*, M) =N_\text{cent}(< m^\mathrm{*}, M)+N_\text{sat}(< m^*, M). \label{cent_sat}
\end{equation}
The central galaxies are modeled as a step function
\begin{equation}
N_\text{cent}(< m^*, M) = \frac{1}{2}  \Biggl[ 1 + \mbox{erf}\Biggl( \frac{\log{M} -  \log{M_\text{min}(m^*) }  }{\sigma(m^*)}
\Biggr)\Biggr], \label{cent}
\end{equation}
while the satellite galaxies are a product of the same step function and a power law
\begin{equation}
N_\text{sat}(< m^*, M) = N_\text{cent}(< m^*, M) \times \Biggl(\frac{M- M_0(m^*)}{ M_1(m^*)}\Biggr)^{\alpha(m^*)}. \label{sat}
\end{equation}
The model described above involves 5 parameters. 
$M_\text{min}$ is the halo mass such that the average number of central galaxies with luminosity above the cut luminosity
is $1/2$, $\sigma$ regulates the efficiency at which the number of galaxies increases from the small to the large halos regime,
$M_0$ is the cutoff mass scale for the satellite galaxies, $M_1$ is a normalization factor and $\alpha$ is the slope of the power-law
that determine the number of galaxies in the highly massive halos regime.
These parameters are not independent: they all depend on the threshold magnitude of the considered catalogue.
In Ref.~\cite{Zehavi:2011} the parameters have been computed for the SDSS galaxy catalogue by considering samples
with different luminosity thresholds. 
In this section we will assume a galaxy sample that follows the same behavior. Nevertheless, the functional
form in the equations \eqref{cent_sat}, \eqref{cent} and \eqref{sat} can be in principle applied to other galaxy catalogues.
Figure \ref{fig:HOD} (left panel) represents the average number of galaxies as a function of the halo mass, for galaxy
samples for different values of the maximum absolute magnitude $M_r$ \footnote{The absolute magnitude is related to the 
apparent magnitude $m^*$ by $M_r = m^* - 5\log_{10}{\frac{d_L(z)}{d_L(z_\text{ref})}} - K(z)$, where $d_L(z)$ is the luminosity distance at a given redshift and $K(z)$ is the k-correction, which corrects the measured magnitude into the one that would be measured in the source's rest frame.}.

By assuming this behaviour for $N_\text{AV}$, we can use the same framework defined in section \ref{Sec:gal} to relate the 
shot-noise, the clustering bias and the magnification bias to the magnitude threshold, at fixed redshift.
In fact, we can compute the comoving number density of galaxies as
\begin{equation}
\bar{n}_\text{gal} (< M_\text{r}) =  \int^{\infty}_{0} n(M, z) N_\text{AV}(< M_\text{r}, M) dM.
\end{equation}
 The shot-noise is the inverse of $\bar{n}_\text{gal} (< M_\text{r})$, while the galaxy and the magnification biases
 are
 \begin{align}
 b_\text{g} (M_\text{r}) &=  \frac{1}{n_\text{g}(z)} \int^{\infty}_0 n(M,z) b(M,z) N_\text{AV}(< m^*, M) dM, \label{HOD_bs} \\
 s_\text{gal}(M_\text{r})& = \frac{\partial \log_{10}(\bar{n}_\text{gal} (< M_\text{r})) }{\partial M_\text{r}}. \label{sgal}
 \end{align} 
Note that since we fixed the redshift, the derivative with respect to the apparent magnitude coincides with the one respect the
absolute magnitude $M_\text{r}$.
The evolution bias can be directly computed from \eqref{fevo_gal}.
In figure \ref{fig:HOD} (right panel) we plot $\bar{n}_\text{gal} (< M_\text{r})$, $ b_\text{g}$ and $s_\text{gal}$ as a function
of the maximum magnitude of the sample. 
We see that samples with a smaller value of the magnitude threshold correspond to higher value of the biases and higher shot-noise
(lower comoving density).
Furthermore, we observe that the magnification bias increases for smaller values of the magnitude cut $M_\text{r}$, reaching 
values up to $\approx 3$.
The evolution bias is also larger for smaller values of $M_\text{r}$, but its value are relatively small in all the range of magnitude threshold.

\begin{figure}[t]
\begin{center}
\includegraphics[width=0.8\textwidth]{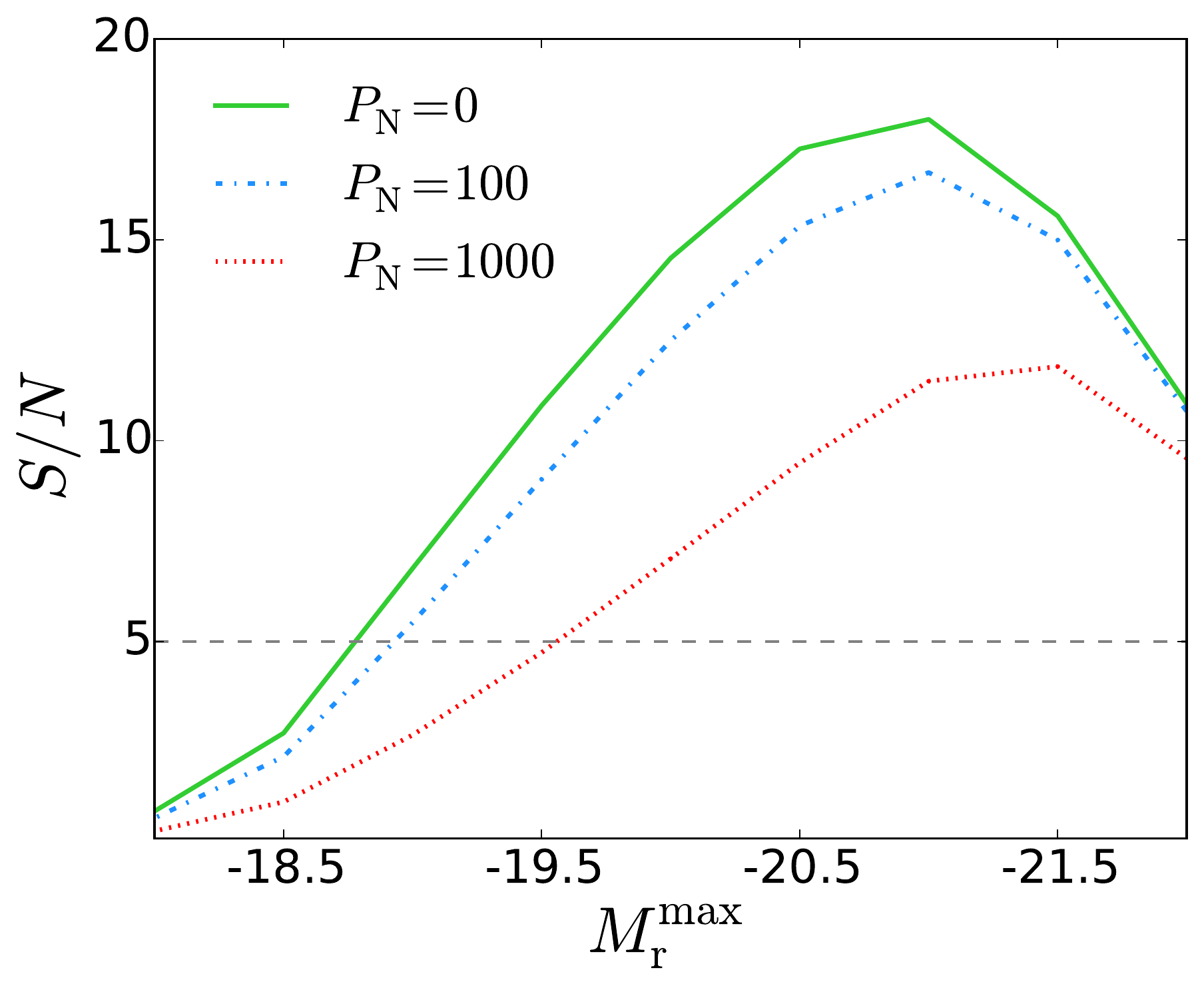}
\caption{Signal-to-noise as a function of the magnitude threshold of the galaxy survey. The galaxies are described by
the HOD model discussed in this section, while the model for the HI is the more conservative model described in the previous section
(see Table \ref{table_HImod}). Different colors and line-styles denote different values of the interferometer noise (the unit is $\text{Mpc}^3/{h}^3$).
The dashed horizontal line denotes a detection threshold of $S/N = 5$.}
\label{HOD_SN}
\end{center}
\end{figure}

Once we have the behaviour of the biases and of the shot-noise in terms of the magnitude cut, we can compute the
correspondent signal to-noise for the dipole from \eqref{SN}. 
The result is plotted in figure \ref{HOD_SN}, where different colors and line-styles refer to different values of the interferometer noise.
We assumed the model 1 (the conservative one) in Table \ref{table_HImod} for the HI, while all the galaxies properties shown in figure \ref{fig:HOD}.
We found that the signal-to-noise is optimized for a certain value of the limiting magnitude, which is not its maximum values (which
corresponds to the minimum shot-noise), because the magnification bias, and therefore the dipole signal, has
larger values if the maximum magnitude threshold is set to be smaller.
 Up to $M_\text{r} \approx - 21$ the signal increases faster than the noise,
while for $M_\text{r} < - 21$ the shot-noise dominates over the signal growth rate.
This result implies that, even if for a galaxy catalogue the expected signal-to-noise for the dipole is below the detection threshold (in figure
\ref{HOD_SN} we set $S/N \ge 5$ for a possible detection), it is possible to properly choose a smaller limiting magnitude for the sample and reject
the galaxies with magnitude above this value, in order to amplify the $S/N$ above the detection threshold.

\section{Conclusions}
\label{Sec:concl}
In this work we studied the relativistic dipole in the cross-correlation function of HI intensity mapping and galaxies.
We model the HI and galaxy parameters (clustering bias, magnification and evolution bias and shot-noise) that affects the signal in the general framework of the halo model. 

We present a signal-to-noise analysis for the relativistic dipole sourced by the Doppler effect of the cross-correlation between the galaxy number counts and the HI brightness temperature.  
Our analysis aims to study the properties of a galaxy population that optimize the signal-to-noise ratio. 
We consider two HI models and in both cases we find that the signal-to-noise is suppressed for highly biased tracers, in despite that the dipole is proportional to the bias difference of the tracers. Therefore galaxies with lower bias are preferred for detecting 
the Doppler dipole in the cross-correlation of HI intensity mapping and galaxies. 
Furthermore the signal appears to be considerably larger for higher values of the magnification bias, therefore galaxy surveys with steeper luminosity
function are favored: we find that in order to have a $S/N > 5$ we need roughly $s \ge 0.5$. 

Finally, we perform a similar signal-to-noise analysis to a luminosity threshold galaxy catalogue. 
We describe the number of galaxies with magnitude below a given threshold using a model based on the Halo Occupation Distribution.
We study the relation between the parameters of the galaxy population and the limiting magnitude. In particular,
we investigate the signal-to-noise of the dipole as a function the magnitude threshold. 
Our results, see figure~\ref{HOD_SN}, indicate that the maximum value for the limiting magnitude of the survey does not correspond to the higher 
signal-to-noise ratio. Indeed, we show how to select an optimal value for the magnitude threshold to maximize the signal-to-noise.
This analysis depends on the specific Halo Occupation Distribution model that we employed, which was built to fit the SDSS data.
Nevertheless, analogue methods can be applied to other galaxy catalogues and can be useful for selecting the optimal galaxy sample to measure the relativistic dipole.

We conclude with some words of caution. First of all both the modelling of the HI and the galaxy populations are based on suitable, physically motivated extensions of the halo model that are expected to be reliable. However, the modelling of HI inside galaxies in the low redshift Universe is crucial and it is not fully explored in this work how observational results of the HI content of galaxies impact on the halo model parameters. In this respect we notice that attempts of modelling the IM signal by incorporating these physical effects into the simple parametric model have been made \cite{villapla16}.
Secondly, here we neglect any modelling of the foreground signal of the IM that we know dominates by several orders of magnitude. Foreground removal techniques are of primary importance in order to fully exploit the cross correlation signal and preliminary results are encouraging (e.g. \cite{alonso15}) also for the IM-galaxy cross-correlation \cite{crosspaco}. The relatively high values of the signal-to-noise ratio of the effect under study is however suggesting that it could still be detected once these caveats are properly modelled.

\acknowledgments
We thank Francesco Montanari and Ruth Durrer for useful discussions.
MV and ED are supported by the cosmoIGM starting grant. FL, EV, MV and ED are supported by the INFN PD51 grant INDARK.
\vspace{1.5cm}

\appendix

\noindent{\LARGE \bf Appendix}

\section{Flat-sky versus full-sky dipole}
\label{Ap_A}

\begin{figure}
\begin{center}
\begin{tikzpicture}[] 
\node[draw,circle] (B) at (0,-2.5) {\textsc{Obs}};
\node[] (A) at (1.7,-1.6) {$-\bn$};
\node[draw,circle] (gal) at (6, 4.0) {{\textsc{Gal}}};
\node[] (zm) at (6, 1.5) {};
\fill (zm) circle [radius=2pt];
\node[] (test) at (0.0, 5.0){};
\node[] (beta) at (8.0,2.83333333333) {};
\draw[dashed, -, thick]  (B) -- (beta);
\draw[dashed, -, thick] (B) -- (zm) node [midway, above, sloped] (TextNode) {$r(z_\text{m})$};;
\node[draw,circle] (h1) at (6, -1.0) {{\textsc{HI}}};
\draw[dashed, -, thick]  (B) -- (gal);
\draw[dashed, -, thick] (B) -- (gal) node [midway, above, sloped] (TextNode) {$r_2$};;
\draw[dashed, -, thick]  (B) -- (h1);
\draw[dashed, -, thick] (B) -- (h1) node [midway, below, sloped] (TextNode) {$r_1$};;
\draw[dashed, -, thick]  (h1) -- (gal);
\draw[->, thick]  (B) -- (2.2, -1.03333333333);
\draw[->, thick]  (h1) -- (6, 0.9);
\draw[<->, thick]  (7, -1.0) -- (7, 4.);
\node[] (dist) at (7.5,1.5) {d};
\node[] (N) at (6.3, 0.0) {$\mathbf{N}$};
\draw[->, thick]  (B) -- (test);
\draw[thick] ([shift=(33:1cm)]6,1.5) arc (33:90:1cm);
\node[] (b) at (2.0,0.4) {$\tilde{\beta}$};
\node[] (b) at (2.0,2.4) {$\alpha$};
\draw[thick] ([shift=(14:3cm)]0,-2.5) arc (14:90:3cm);
\draw[thick] ([shift=(14:3.1cm)]0,-2.5) arc (14:90:3.1cm);
\draw[thick] ([shift=(47:5cm)]0,-2.5) arc (47:90:5cm);\node[] (b) at (6.6,2.7) {$\beta$};
\node[] (A1) at (1.5,-2.4) {$-\bn_1$};
\node[] (A2) at (0.8,-1.0) {$-\bn_2$};
\draw[->, thick]  (B) -- (1.8, -0.55);
\draw[->, thick]  (B) -- (2.5, -1.875);
\end{tikzpicture}
\end{center}  
\caption {Coordinate system for the full-sky relativistic dipole.}
\label{fig_apA}
\end{figure}
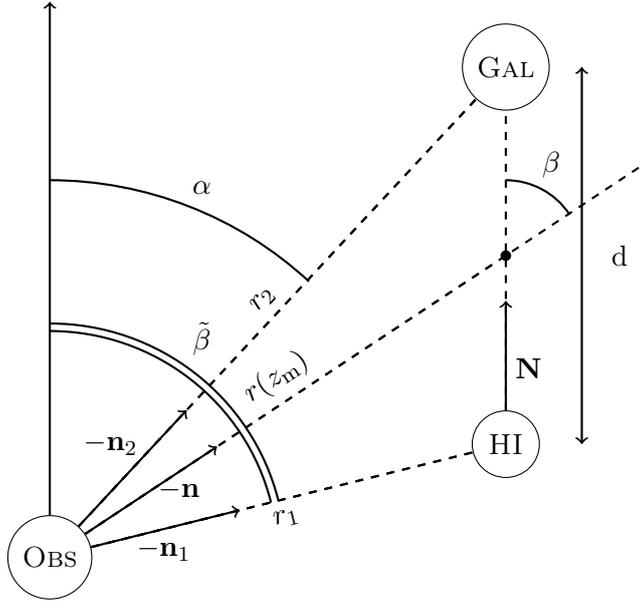

In this Appendix we test the validity of the flat-sky approximation by comparing the flat-sky
and the full-sky relativistic dipole at redshift $z_{m} = 0.15$.
This comparison is crucial to test in which regime we can apply the flat-sky approximation and therefore which are the maximum scales
that we can model in this framework. In particular, we test the dipole expansion in terms of $d/r$ which, to lowest order, leads to the wide-angle correction, defined in Eq.~\eqref{wa_corr}, that we adopted in our analysis.

The full-sky correlation function for the cross-correlation of two populations of galaxies has been studied in detail in
the literature, both in the newtonian approximation \cite{Szalay:1997cc, Szapudi:2004gh, Papai:2008bd} and beyond \cite{Raccanelli:2010hk, Jeong:2011as, samushia2012, Bertacca:2012, Bonvin:2013, Raccanelli:2013, YooSeljak,  Tansella:2017rpi}.
Here we will use the same notation as \cite{Bonvin:2013}, where the full-sky relativistic correlation function is computed for 
two galaxy populations, including the gravitational redshift and Doppler effects, which are the quantity relevant for this work
(see the Appendix B in \cite{Bonvin:2013} for a detailed derivation of the relativistic part, i.e.~the cross-correlation between the density plus redshift space distortion and the Doppler terms in the galaxy number counts).
We will apply that expression to the cross-correlation between HI and galaxies.
We will fix the values of the bias to $b_\text{HI} = 0.6$ and $b_\text{gal} = 1.0$, and we will compare 
three values of the galaxy magnification bias. 

We will recast the expression for the full-sky correlation function in the way described below.
The relativistic correlation function, in the coordinate system defined in section \ref{sub2.3} (see also
figure \ref{fig_apA}), can be written in terms of some coefficients $R_i = R_i (r_1, r_2, d)$

\bea
\xi^{\text{HI}, \text{gal}}(r, d, \mu) &= &R_1 \cos{(\alpha)} + R_2  \cos{(\tilde{\beta)}}  + R_3 \cos{(\alpha)}\cos{(2\tilde{\beta})} + R_4 
 \cos{(\tilde{\beta)}} \cos{(2\alpha)}  \label{full_sky_cf} \notag \\
&&+ R_5 \sin{(\alpha)}\sin{(2\tilde{\beta})} + R_6 \sin{(\tilde{\beta})}\sin{(2\alpha)}, 
\eea
where $\alpha$ is defined as $\cos{(\alpha)} =  -\mathbf{n_2} \cdot \mathbf{N}$ and
$\cos{(\tilde{\beta})} = - \mathbf{n_1} \cdot \mathbf{N}$ (see figure \ref{fig_apA}), while the coefficients $R_i$
are 
\begin{align}
R_1 = & C_\text{gal}(r_2) \frac{D_1(r_2)D_1(r_1)}{D_1^2(r)} f(r_2) \Biggl[ \Biggl( b_\text{HI} +\frac{2}{5}f(r_1) \Biggr)\nu_1 -\frac{1}{10} f(r_1) \nu_3\Biggr], \notag \\
R_2 = &-C_\text{HI} (r_1) \frac{D_1(r_2)D_1(r_1)}{D_1^2(r)} f(r_1) \Biggl[ \Biggl( b_\text{gal} +\frac{2}{5}f(r_2) \Biggr)\nu_1 -\frac{1}{10} f(r_2) \nu_3\Biggr], \notag \\
R_3 = & C_\text{gal} (r_2)\frac{D_1(r_2)D_1(r_1)}{D_1^2(r)}  f(r_1)f(r_2) \frac{1}{5}\Biggl[\nu_1 -\frac{3}{2} \nu_3\Biggr],\notag \\
R_4 = & -C_\text{HI} (r_1) \frac{D_1(r_2)D_1(r_1)}{D_1^2(r)} f(r_1)f(r_2) \frac{1}{5}\Biggl[\nu_1 -\frac{3}{2} \nu_3\Biggr], \notag \\
R_5 = & C_\text{gal} (r_2) \frac{D_1(r_2)D_1(r_1)}{D_1^2(r)}  f(r_1)f(r_2) \frac{1}{5}\Biggl[\nu_1 + \nu_3\Biggr],\notag \\
R_6 = & -C_\text{HI} (r_1) \frac{D_1(r_2)D_1(r_1)}{D_1^2(r)}  f(r_1)f(r_2) \frac{1}{5}\Biggl[\nu_1 + \nu_3\Biggr], 
\end{align}
where $C_\text{gal}$ and $C_\text{HI}$ are defined in \eqref{coeff}, $D_1$ denotes the linear growth factor, and $\nu_{\ell=1,3}$ are defined through
\begin{equation}
\nu_{\ell}(d) = \int \frac{k^2 dk}{2\pi^2}\Biggl(\frac{\mathcal{H}}{k}\Biggr) P(k, z_{m}) j_{\ell} (k\,d).
\end{equation}
All the quantities involved in the computation of the correlation function in \eqref{full_sky_cf} can be expressed as a function of $r$, $d$ and $\mu$ by using the following simple geometric relations
\bea \label{geom_rel}
r_1 = \frac{1}{2}\sqrt{d^2 + 4r^2 - 4 d \mu r}\, , \qquad  && \qquad  r_2 = \frac{1}{2}\sqrt{d^2 + 4r^2  + 4 d \mu r}\,  , \nonumber \\
 \cos{(\tilde{\beta})} = \frac{-d+2 r\mu}{\sqrt{4 r^2 + d^2 -4 d r\mu}} \,  , \qquad &&   \qquad  \cos{(\alpha)} = \frac{d+2 r\mu}{\sqrt{4 r^2 + d^2 +4 d r\mu}}\,  .
\eea
The dipole can then be computed as
\begin{equation}
\xi^\text{full sky}_{1} = \frac{3}{2}\int_{-1}^1 \xi^{\text{HI}, \text{gal}} (r, d, \mu)  L_1(\mu) d \mu. \label{full_sky_multipoles}
\end{equation}
We see the full-sky correlation function depends in a not trivial way from the angular coordinate $\mu$. Therefore, the angular integral in \eqref{full_sky_multipoles} cannot
be performed analytically, as in the flat-sky approximation, but it needs to be solved numerically. 

A similar computation can be done for the wide-angle corrections.
The full sky expression can be written in terms of some coefficients $S_i = S_i (r_1, r_2, d)$ \cite{Bonvin:2013}
\begin{equation}
\xi^\text{WA}(r, d, \mu) = S_1 + S_2  \cos{(2 \tilde{\beta)}}  + S_3 \cos{(2 \alpha)} +
S_4 \cos{(2 \alpha)}  \cos{(2 \tilde{\beta)}}   + S_5 \sin{(2\tilde{\beta})}\sin{(2\alpha)}, 
\end{equation}

\begin{align}
S_1 = &\frac{D_1(r_2)D_1(r_1)}{D_1^2(r)} f(r_2) \Biggl[ \Biggl( b_\text{HI} b_\text{gal} + \frac{b_\text{HI}}{3}f(r)
+ \frac{b_\text{gal}}{3}f(r)  +\frac{2}{15} f^2(r)\Biggr)\mu_0, \notag \\
&-\frac{1}{3} \Biggl(  \frac{b_\text{HI}}{2}f(r) +  \frac{b_\text{gal}}{2}f(r) +\frac{2}{7}  f^2(r)  \Biggr) \mu_2
+\frac{3}{140}  f^2(r) \mu_4 \Biggr], \notag \\
S_2 = &- \frac{D_1(r_2)D_1(r_1)}{D_1^2(r)} \Biggl[ \Biggl(\frac{b_\text{gal}}{2}f(r) +\frac{3}{14}f^2(r) \Biggr)\mu_2 
-\frac{1}{28} f^2(r) \mu_4 \Biggr],  \notag \\
S_3 = & -\frac{D_1(r_2)D_1(r_1)}{D_1^2(r)} \Biggl[ \Biggl(\frac{b_\text{HI}}{2}f(r) +\frac{3}{14}f^2(r) \Biggr)\mu_2 
-\frac{1}{28} f^2(r) \mu_4 \Biggr],\notag \\
S_4 = & \frac{D_1(r_2)D_1(r_1)}{D_1^2(r)}  f(r)^2 \frac{1}{5} \Biggl[\mu_0 - \frac{1}{21}\mu_2 +\frac{19}{140}\mu_4 \Biggr], \notag \\
S_5 = & \frac{D_1(r_2)D_1(r_1)}{D_1^2(r)}  f(r)^2 \frac{1}{5} \Biggl[\mu_0 - \frac{1}{21}\mu_2 -\frac{4}{35}\mu_4 \Biggr],
\end{align}
where  $\mu_{\ell=0, 2, 4}$ are defined by
\begin{equation}
\mu_{\ell}(d) = \int \frac{k^2 dk}{2\pi^2} P(k, z_{m}) j_{\ell} (k\,d).
\end{equation}
The dipole of the wide-angle effect, in full sky, can be computed as
\begin{equation}
\xi^\text{WA, full sky}_{1} = \frac{3}{2}\int_{-1}^1 \xi^\text{WA} (r, d, \mu)  L_1(\mu) d \mu. \label{full_sky_WA}
\end{equation}

\begin{figure}[t]
\centering
    \begin{subfigure}[b]{0.495\textwidth}
     \begin{center}
\includegraphics[width=1.0\textwidth]{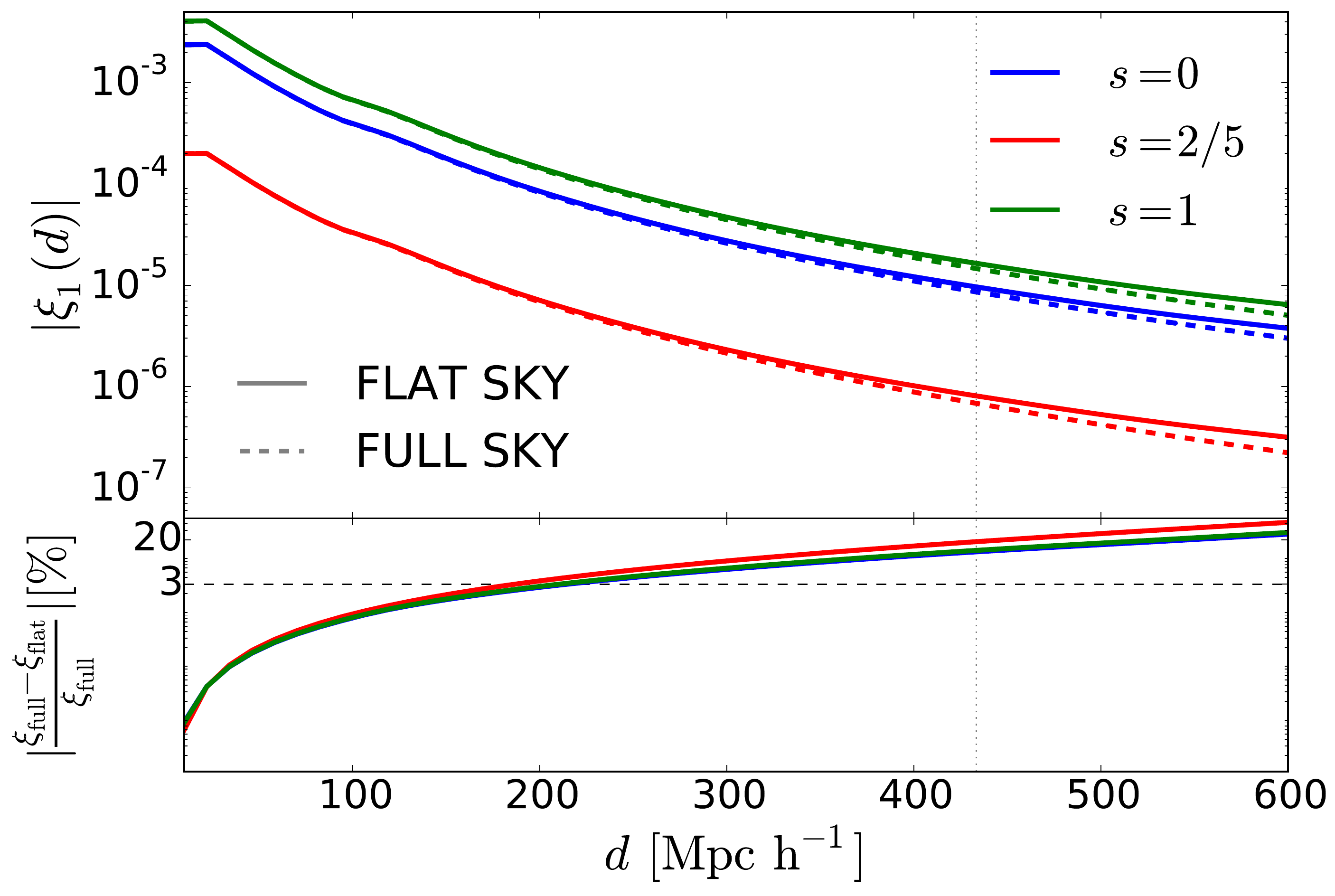}
\caption{Doppler dipole.}
\label{dip_full}
\end{center}
    \end{subfigure}
    \begin{subfigure}[b]{0.495\textwidth}
     \begin{center}
\includegraphics[width=1.0\textwidth]{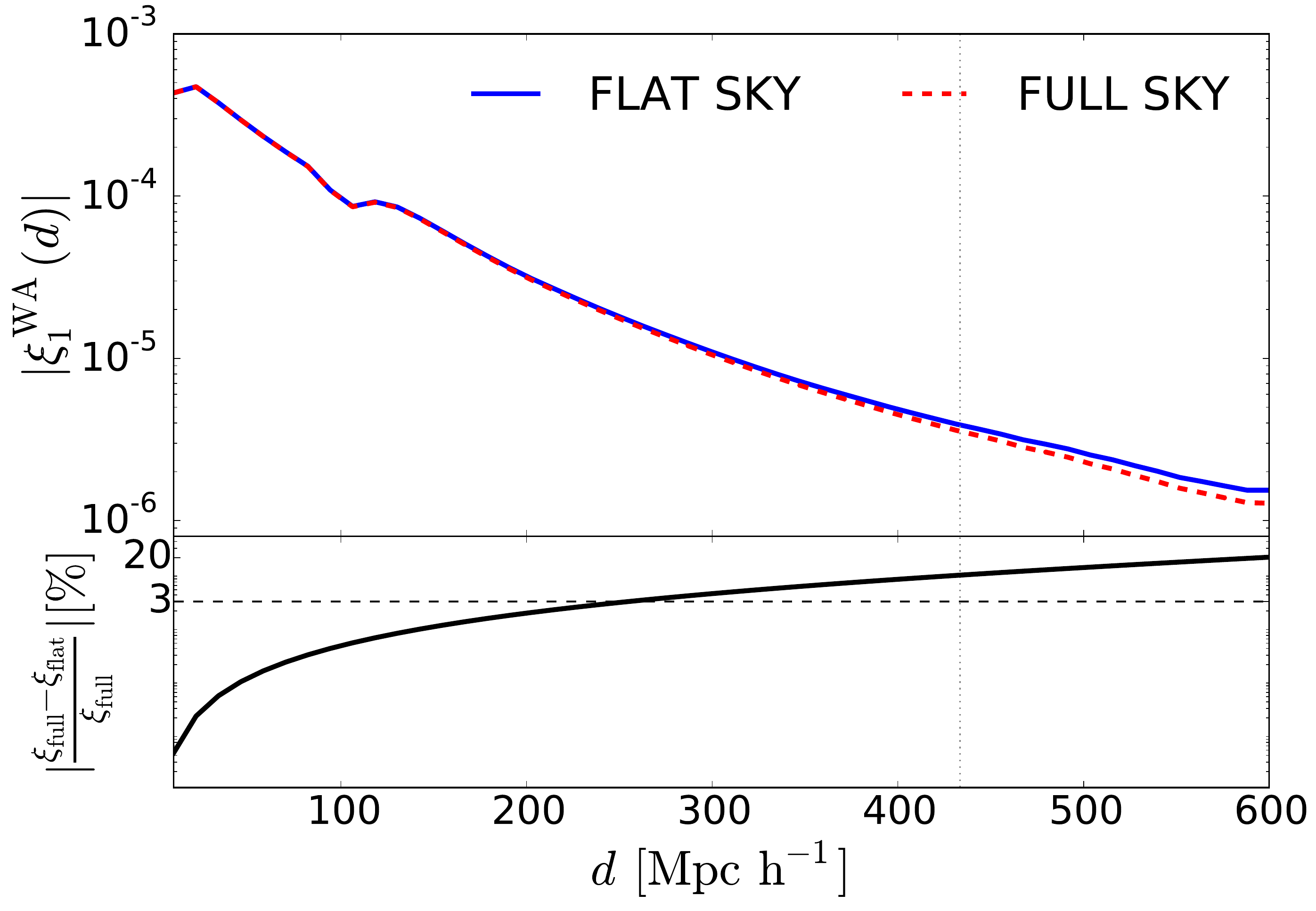}
\caption{Wide-angle dipole.}
\label{WA_full}
\end{center}
    \end{subfigure}
    \caption{Comparison between the dipole of the cross-correlation HI-galaxies (right panels)  and the wide-angle dipole (left panels) 
    computed within the flat-sky approximation (continuous line) and the full-sky dipole signal (dashed line). 
Top panels: absolute value of the dipole.
Bottom panels: relative difference between the full signal and the flat-sky approximation. The dashed line denotes a difference of relative 
difference of $3\%$.
Different colors denote different values of the galaxy magnification bias. The dotted vertical lines denotes the distance $d = r$.
}
    \label{fig_fullsky}
\end{figure}

\begin{figure}[t]
\begin{center}
\includegraphics[width=0.8\textwidth]{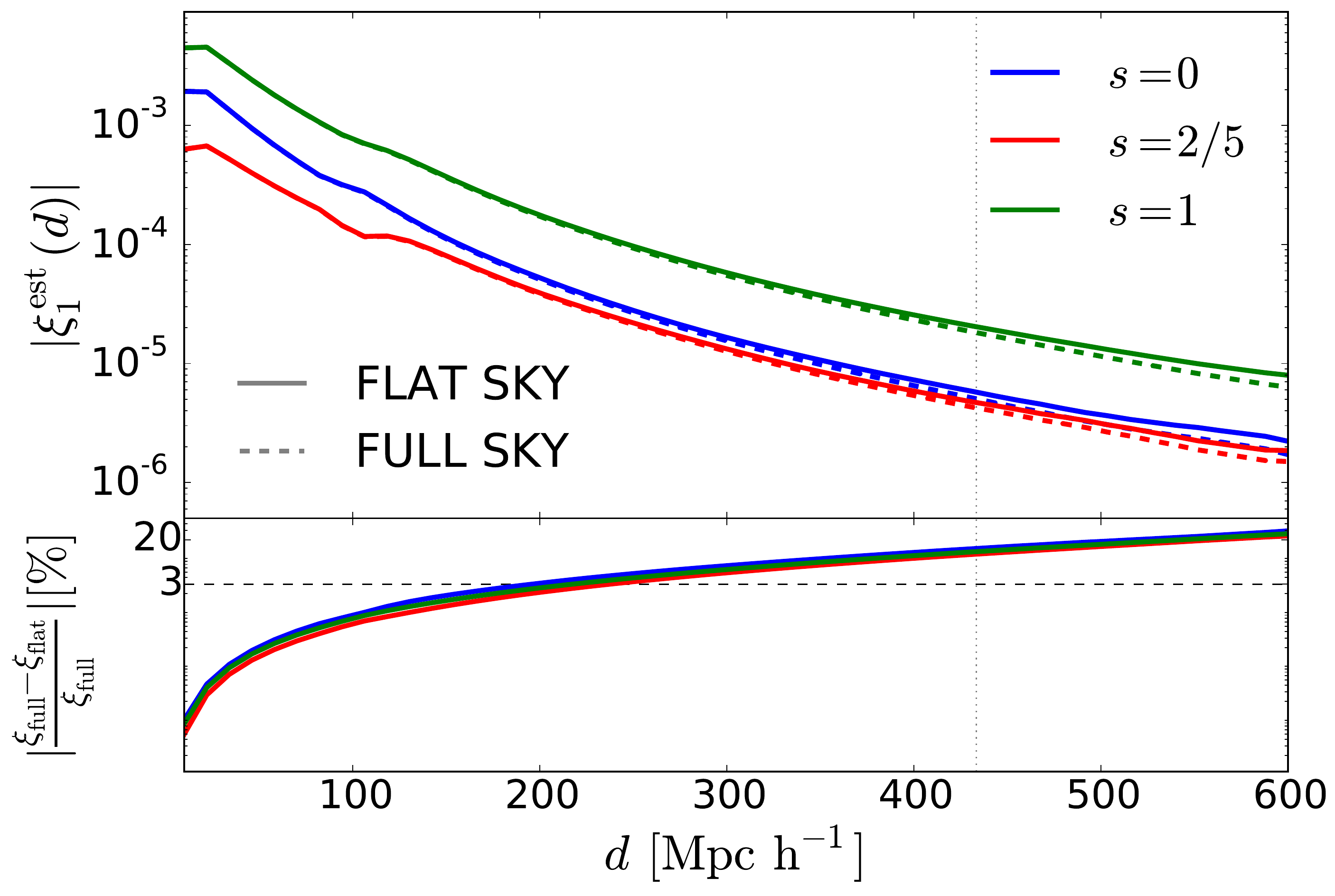}
\caption{Comparison between the estimator of the cross-correlation HI-galaxies computed within the flat-sky approximation (continuous line) and the full-sky dipole signal (dashed line).
Top panel: absolute value of the estimator. Bottom panel: relative difference between the full signal and the approximated signal.
Different colors denote different values of the galaxy magnification bias.
}
\label{est_fullsky}
\end{center}
\end{figure}

In figure \ref{fig_fullsky} we compare the full-sky dipole (dashed lines) and the corresponding dipole in the flat-sky approximation (continuous lines).
Figure \ref{dip_full}  represents the Doppler dipole, computed from \eqref{full_sky_multipoles} (full-sky) and \eqref{multipoles} (flat-sky),
while figure \ref{WA_full} shows the wide-angle dipole, computed from \eqref{full_sky_WA} (full-sky) and \eqref{wa0} (flat-sky). 
Different colors refer to different values of the magnification bias.
In the top panel we plot the absolute value of the dipole. At small scale, the flat-sky approximation fairly reproduces the full-sky signal, while
the full-sky dipoles significantly departs from the one in flat-sky at large scales. Note that the full-sky signal always results to be smaller, in absolute value, than the approximated one. Indeed two sources appear closer in the flat-sky limit and therefore they seems to be more correlated.
In the bottom panel we plot the relative difference between the full-sky and the flat-sky dipole, in percentage.
The black dashed line denotes a 3\% difference between the two quantities, which we set as a threshold for the flat-sky approximation to be valid. 
We see that for scales larger than roughly $200 \text{Mpc}/\text{h}$ the relative difference is beyond the threshold value and that it rapidly increases at larger scales. Nevertheless the amount of physical information at these scales is irrelevant as we see from figure \ref{fig:dmin_dmax}.

In figure \ref{est_fullsky} we show the difference between the estimator defined in \eqref{wa_corr}, which is unbiased from wide-angle effect,
in full-sky and the one in flat-sky, at lower order in the $d/r$ expansion.
We see that at scales $d \le 200 \, \text{Mpc}/\text{h}$ the approximated dipole fairly reproduces the full-sky quantity.
Therefore, we set the maximum scale of our analysis to be $d_\text{max} = 200 \,\text{Mpc}/\text{h}$.

\section{Contribution to the wide-angle correction to the covariance}
\label{Ap_B}

The estimator we used in the signal-to-noise analysis, which is unbiased from wide-angle effects (at least at the leading order in the $d/r$ expansion),
involved the measurement of both the dipole of the cross-correlation of galaxies and HI and the quadrupole of the autocorrelations of both 
tracers 
\begin{equation}
\hat\xi_1(d, r) = \hat\xi_1(d, z_1) - \frac{3}{10} \left(\hat\xi^{gal}_2(d, z_1) - \hat\xi^\text{HI}_2(d, z_1)\right)\frac{d}{r}. \label{wa_corr_A}
\end{equation}
We split the covariance for the estimator in Eq.~\eqref{wa_corr_A} in two contributions \footnote{We assume the redshift to be constant 
and we drop the redshift dependence from our notation},
\begin{equation}
\mbox{COV}^\text{est}(d_1, d_2) = \mbox{COV}^\text{dip}(d_1, d_2) + \mbox{COV}^\text{WA}(d_1, d_2),
\end{equation}
where $ \mbox{COV}^\text{dip}(d_1, d_2) = \langle\hat\xi_1(d_1) \hat\xi_1(d_2) \rangle$ is the dipole contribution to the estimator it is given by Eq.~\eqref{COV},
while $\mbox{COV}^\text{WA}(d_1, d_2)$ is the contributions of the wide-angle correction,
\bea
\mbox{COV}^\text{WA}(d_1, d_2) = & \left(\frac{3}{10r}\right)^2 d_1\, d_2  \left( \mbox{COV}^\text{gal, gal}_{2,2} + 
\mbox{COV}^\text{HI, HI}_{2,2}  + 2  \, \mbox{COV}^\text{gal, HI}_{2,2} \right) \nonumber \\
 & - \frac{3}{10} \frac{d_1+d_2}{r}\left( \mbox{COV}^\text{Dip, gal}_{1,2} - \mbox{COV}^\text{Dip, HI}_{1,2} \right), \label{cross}
\eea
where 
\begin{align}
\mbox{COV}^\text{gal, gal}_{2,2}(d_1, d_2) &= \mean{\hat\xi^{gal}_2(d_1) \hat\xi^{gal}_2(d_2)}  -  \mean{\hat\xi^{gal}_2(d_1)} \mean{\hat\xi^{gal}_2(d_2)}, \notag \\
\mbox{COV}^\text{HI, HI}_{2,2} (d_1, d_2)&=  \mean{\hat \xi^{HI}_2(d_1) \hat\xi^{HI}_2(d_2)} -  \mean{\hat \xi^{HI}_2(d_1)} \mean{\hat\xi^{HI}_2(d_2)}, \notag \\
\mbox{COV}^\text{gal, HI}_{2,2}(d_1, d_2) &=  \mean{\hat\xi^{gal}_2(d_1) \hat\xi^{HI}_2(d_2)} -  \mean{\hat\xi^{gal}_2(d_1)}\mean{\hat\xi^{HI}_2(d_2)}, \notag \\
\mbox{COV}^\text{Dip, gal}_{1,2}(d_1, d_2) &= \mean{\hat\xi_1(d_1) \hat\xi^{gal}_2(d_2)} - \mean{\hat\xi_1(d_1)} \mean{  \hat\xi^{gal}_2(d_2)}, \notag \\
\mbox{COV}^\text{Dip, HI}_{1,2}(d_1, d_2) & = \mean{ \hat\xi_1(d_1) \hat\xi^{HI}_2(d_2)}  - \mean{ \hat\xi_1(d_1)} \mean{\hat\xi^{HI}_2(d_2)}.
\end{align}
The first line in \eqref{cross} is the autocorrelation of the wide-angle contribution to the dipole, and the second one is the cross-correlation
of the dipole with the wide-angle correction.
Each term of the wide-angle contribution to the covariance can be computed similarly to the dipole contribution by applying 
the general formula in \cite{Hall:2016}.

The cosmic variance contribution to the autocorrelation can be written as
\begin{equation}
\mbox{COV}^\text{WA, auto}_\text{CVCV} =  \Biggl(\frac{3}{10r}\Biggr)^2
\frac{50 d_1 d_2}{V} \int \frac{k^2 dk}{2\pi^2} j_{2}(k d_1) j_{2}(k d_2) \sum_{L_1, L_2} G_{22}^{L1L2} [P^\text{gal}_{L1} P^\text{gal}_{L2} + 
P^\text{HI}_{L1} P^\text{HI}_{L2} - 2 P^\text{gal}_{L1} P^\text{HI}_{L2}]; \label{auto1}
\end{equation}
the mixed cosmic variance -- noise contribution to the autocorrelation, including both shot-noise and interferometer noise, is given by
\begin{equation}
\mbox{COV}^\text{WA, auto}_\text{CVNoise} =  \Biggl(\frac{3}{10r}\Biggr)^2
\frac{50 d_1 d_2}{V}  \int \frac{k^2 dk}{2\pi^2} j_{2}(k d_1) j_{2}(k d_2) \sum_{L} G_{22}^{0L} [1/n_\text{gal} P^\text{gal}_{L}+ 
(N_\text{HI} + P_\text{N})P^\text{HI}_{L}]; \label{auto2}
\end{equation}
while the noise contribution to the covariance is
\begin{equation}
\mbox{COV}^\text{WA, auto}_\text{NoiseNoise} =  \Biggl(\frac{3}{10r}\Biggr)^2
\frac{10}{4\pi L_p V} \delta_{d_1, d_2}\Biggl[(N_\text{HI} + P_\text{N})^2 +\frac{1}{n^2_\text{gal}}\Biggr].
\end{equation}
The coefficients $G_{22}^{L1L2}$ in \eqref{auto1} and \eqref{auto2} are defined in terms of the Wigner $3j$ symbols
\begin{equation}
G_{22}^{L1L2} = \sum_L (2L + 1) \tj{2}{2}{L}{0}{0}{0}^2  \tj{L_1}{L_2}{L}{0}{0}{0} ^2.
\end{equation}
The cosmic variance contribution to the cross-correlation term in \eqref{cross} can be written as
\begin{equation}
\mbox{COV}^\text{WA, cross}_\text{CVCV} =  \Biggl(\frac{3}{10} \frac{d_1+d_2}{r}\Biggr)
\frac{30}{V} \int \frac{k^2 dk}{2\pi^2} j_{1}(k d_1) j_{2}(k d_2) \sum_{L_1, L_2} G_{12}^{L1L2} P_{L1} [P^\text{gal}_{L2} -P^\text{HI}_{L2}]; 
\label{cross1}
\end{equation}
the cosmic variance x noise contribution to the cross-correlation is given by
\begin{equation}
\mbox{COV}^\text{WA, cross}_\text{CVNoise} =  \Biggl(\frac{3}{10} \frac{d_1+d_2}{r}\Biggr)
\frac{30}{V} \int \frac{k^2 dk}{2\pi^2} j_{1}(k d_1) j_{2}(k d_2) \sum_{L} \Biggl(\frac{1}{n_\text{gal}} - N_\text{HI} - P_\text{N}\Biggr)G_{12}^{0 L} P_{L}.  \label{cross2}
\end{equation}
The coefficients $G_{12}^{L1L2}$ are defined similarly to the coefficients involved in the autocorrelation term,
\begin{equation}
G_{12}^{L1L2} = \sum_L (2L + 1) \tj{1}{2}{L}{0}{0}{0}^2  \tj{L_1}{L_2}{L}{0}{0}{0} ^2.
\end{equation}
The purely noise contribution to the cross-correlation term vanishes.

\bibliographystyle{JHEP}
\bibliography{mybib}

\end{document}